\documentclass[prc,twocolumn,showpacs,superscriptaddress,preprintnumbers,amsmath,amssymb,nofootinbib]{revtex4-1}
\usepackage{bm}
\usepackage{dcolumn}
\usepackage{multirow}
\usepackage{ifpdf}
\usepackage{url}

\ifpdf
\usepackage{graphicx} 
\usepackage{hyperref}
\else 
\usepackage[dvipdfmx]{graphicx} 
\usepackage[dvipdfmx]{hyperref}
\fi
\usepackage{color} 

\newcommand{\refbra}[1]{(\ref{#1})}

\newcommand{\dis}{\displaystyle}

\newcommand{\dndeta}[1][flat]
{
    \ifthenelse{\equal{#1}{flat}}{$\left< dN_{\mathrm{ch}}/d\eta\right>$}{}
    \ifthenelse{\equal{#1}{vertical}}{$\left< \cfrac{dN_{\mathrm{ch}}}{d\eta\right>}$}{}
}

\newcommand{\snn}[1][nucleus]
{
    \ifthenelse{\equal{#1}{nucleus}}{$\sqrt{s_{\mathrm{NN}}}$}{}
    \ifthenelse{\equal{#1}{proton}}{$\sqrt{s}$}{}
}

\newcommand{\vtwtw}{v_{2}\{2\}}

\newcommand\pythia{\textsc{Pythia}}
\newcommand\ISthreeD{\textsc{iS3D}}
\newcommand\jam{\textsc{JAM}}

\hypersetup{
colorlinks=true,
linkcolor=blue,
citecolor=blue,
urlcolor=blue
}

\begin{document}

\title{
Interplay between core and corona components \\
in high-energy nuclear collisions
}
\author{Yuuka Kanakubo}
\email{y-kanakubo-75t@eagle.sophia.ac.jp}
\affiliation{%
Department of Physics, Sophia University, Tokyo 102-8554, Japan
}

\author{Yasuki Tachibana}
\email{ytachibana@aiu.ac.jp}
\affiliation{
Akita International University, Yuwa, Akita-city 010-1292, Japan
}
\author{Tetsufumi Hirano}
\email{hirano@sophia.ac.jp}
\affiliation{%
Department of Physics, Sophia University, Tokyo 102-8554, Japan
}

\date{\today}

\begin{abstract}
We establish the updated version of dynamical core--corona initialization framework (DCCI2) as a unified description from small to large colliding systems and from low to high transverse momentum ($p_T$) regions.
Using DCCI2, we investigate effects of interplay between locally equilibrated and non-equilibrated systems, in other words, core and corona components in high-energy nuclear collisions.
Given experimental multiplicity distributions and yield ratios of $\Omega$ baryons to charged pions as inputs, we extract the fraction of core and corona components in p+p collisions at $\sqrt{s}=7$ TeV and Pb+Pb collisions at $\sqrt{s_{NN}}=2.76$ TeV.
We find core contribution overtakes corona contribution as increasing multiplicity above $\langle dN_{\mathrm{ch}}/d\eta \rangle_{|\eta|<0.5} \sim 18$ regardless of the collision system or energy.
We also see that
the core contribution exceeds the corona contribution only in 0.0-0.95\%  multiplicity class in p+p collisions.
Notably, there is a small enhancement of corona contribution with $\sim20$\% below $p_T\sim 1$ GeV even in minimum bias Pb+Pb collisions.
We find that the corona contribution at low $p_T$ gives $\sim 15$-$38$ $\%$ correction on $v_2\{2\}$ at $N_{\mathrm{ch}}\lesssim 370$.
This raises a problem in conventional hydrodynamic analyses in which low $p_T$ soft hadrons originate solely from core components.
We finally scrutinize the roles of string fragmentation and the longitudinal expansion in the transverse energy per unit rapidity, which is crucial in initial conditions for hydrodynamics from event generators based on string models.
\end{abstract}

\pacs{25.75.-q, 12.38.Mh, 25.75.Ld, 24.10.Nz}

\maketitle

\section{Introduction}
\label{sec:intro}
Quark-gluon plasma (QGP) is a state of thermalized and chemically equilibrated matter consisting of quarks and gluons deconfined from hadrons at extremely high temperature and density.
High-energy nuclear collision experiments in the Relativistic Heavy Ion Collider (RHIC) at Brookhaven National Laboratory 
and at the Large Hadron Collider (LHC) in CERN
provide opportunities to explore properties of the extreme state.

Relativistic hydrodynamics is proven to successfully describe experimental data of relativistic heavy-ion collisions since the first discovery of hydrodynamic behavior of the QGP in the early 2000s \cite{Heinz:2001xi,Gyulassy:2004zy,Shuryak:2004cy,Hirano:2005wx}.
Since final observables reflect all the history of the reaction, it is of significant importance to model each stage of the reaction and to integrate these modules as a whole in a consistent way towards a further comprehensive understanding of the QGP \cite{Hirano:2012kj}. In particular, modeling of initial pre-equilibrium and final decoupling stages is needed in addition to a relativistic hydrodynamic model as a framework to describe transient states. 
Notably, there are some attempts to constrain transport coefficients of the QGP by using state-of-the-art dynamical models based on relativistic hydrodynamics \cite{Bernhard:2019bmu,Nijs:2020ors,Everett:2020xug}.
As one sees from this, the QGP study is in the middle of a transition to precision science. 

Despite the great success of dynamical models based on relativistic hydrodynamics in describing a vast body of experimental data, it poses some open issues for a comprehensive description of the whole reaction in high-energy nuclear collisions.
One of the major issues is an initial condition of relativistic hydrodynamic equations which does not respect the total energy of the colliding systems.
Initial conditions have been parametrized and put to reproduce centrality dependence of multiplicity or pseudorapidity distributions in a conventional hydrodynamic approach.
As a result, the total energy of the initial hydrodynamic fields does not exactly match the collision energy of the system. 
Even when some outputs from event generators with a given collision system and energy are utilized for initial conditions in hydrodynamic models, an additional scale parameter is commonly introduced to adjust the model outputs of multiplicity. 
One might not think it is necessary for the energy of the initial hydrodynamic fields to be the same as the total energy of the system.
This is exactly a starting point of our discussion in a series of papers \cite{Okai:2017ofp,Kanakubo:2018vkl,Kanakubo:2019ogh}: The relativistic hydrodynamics merely describes a part of system, namely, matter in local equilibrium, while other parts of the system such as propagating jets and matter out of equilibrium are to be described at the same time.

First attempts of simultaneous description of both the QGP fluids in equilibrium and the energetic partons out of equilibrium had been made in Refs.~\cite{Hirano:2002sc,Hirano:2003hq,Hirano:2003yp,Hirano:2003pw,Hirano:2004en} \footnote{Note that the very first study to utilize the hydrodynamic solutions in quantitative analysis of parton energy loss was done in Ref.~\cite{Gyulassy:2001kr}.}.
Initial conditions in those studies were still either parametrized via an optical Glauber model \cite{Glauber:2006gd} or taken from a saturation model \cite{Kharzeev:2000ph,Kharzeev:2001gp,Kharzeev:2001yq,Kharzeev:2002ei} so as to reproduce yields of low $p_T$ hadrons, while hard partons which undergo energy loss during traversing QGP fluids were supplemented to successfully reproduce the hadron spectra from low to high $p_{T}$ regions \cite{Hirano:2002sc,Hirano:2003yp,Hirano:2004en}.
It was found that an intriguing interplay between soft and hard components brought ones to interpretation of the proton yield anomaly in $p_{T}$ spectra \cite{Hirano:2003yp}.
However, the model lacked back reactions from quenching partons to the QGP fluids and a contribution from fragmentation was cut in low $p_T$ regions, both of which obviously violate the energy-momentum conservation law.

Medium responses to propagating energetic partons have been modeled within hydrodynamic equations with source terms by assuming the instantaneous equilibration of the deposited energy and momentum from partons \cite{Stoecker:2004qu,CasalderreySolana:2004qm,Satarov:2005mv,Renk:2005si,Chaudhuri:2005vc,Chaudhuri:2006qk,Chaudhuri:2007vc,Betz:2010qh,Tachibana:2014lja,Tachibana:2015qxa,Tachibana:2017syd,Chen:2017zte,Chang:2019sae,Tachibana:2020mtb,Zhao:2021vmu}.
Within this approach, the sum of the energy and momentum of fluids and those of traversing partons is conserved as a whole. However, it is not clear how to divide the initial system just after the collision into soft (fluids in equilibrium) and hard (partons out of equilibrium) parts.

To remedy this issue, a dynamical initialization model \cite{Okai:2017ofp} was proposed to describe the dynamics of gradually forming QGP fluids phenomenologically \footnote{Dynamical initialization is essential in describing the formation of fluids in lower collision energies in which secondary hadrons are gradually produced in finite time duration due to insufficient Lorentz contraction of colliding nuclei \cite{Shen:2017bsr,Akamatsu:2018olk}.
}.
In contrast to the conventional hydrodynamic models in which initial conditions are put at a fixed initial time, the QGP fluids are generated locally in time and space in the dynamical initialization framework.
Under this framework, all the input of energy-momentum of QGP fluids is the one of partons produced just after the nuclear collisions. 
Starting with vacuum, energy-momentum of the QGP fluids is dynamically generated by solving hydrodynamic equations with source terms. 
Consequently, fluids in local equilibrium are generated from the initial partons by depositing the energy and the momentum and surviving partons are considered to remain out of equilibrium.
When initial partons are taken, \textit{e.g.}, from event generators, the total energy keeps its value of the colliding two nuclei all the way through the dynamical initialization. 
Although we successfully separated matter in local equilibrium from initially produced partons in the dynamical initialization framework \cite{Okai:2017ofp}, the fluidization scheme was too simple and phenomenological to describe the transverse momentum spectra and the particle ratios.
Then, we introduced the core--corona picture into the dynamical initialization.

The conventional core--corona picture was proposed to explain centrality dependence of strange hadron yield ratios~\cite{Werner:2007bf}. 
As multiplicity increases, the high-density region, in which the matter is mostly thermalized, is supposed to become larger. 
As a result, the final hadron yields become dominated by the hadrons from thermalized matter rather than non-thermalized matter created in low-density regions. 
The former component is referred to as core, while the latter one is referred to as corona. 

A Monte-Carlo event generator, EPOS~\cite{Pierog:2013ria,Werner:2018yad} is widely accepted for its implementation of the core--corona picture. In the latest study in Refs.~\cite{Werner:2013tya,Werner:2019aqa}, string segments produced in a collision are separated sharply into the core and the corona components depending on their density and transverse momentum at a fixed time. Low-$p_T$ string segments in dense regions are fully converted into the thermalized medium fluid, while string segments with high momentum or dilute regions are directly hadronized.

On the other hand, we model the dynamical aspects of the core--corona separation introducing the particle density dependence of the dynamical initialization scheme, which is called the \textit{dynamical core--corona initialization (DCCI)}  \cite{Kanakubo:2018vkl,Kanakubo:2019ogh} \footnote{In fact, this idea was first implemented in Ref.~\cite{Akamatsu:2018olk} to describe excitation functions of particle ratios at lower collision energies. However, it was applied to the secondary produced hadrons rather than partons.}.
One of the key features of the DCCI is to deal with dynamics of the core (equilibrium) and the corona (non-equilibrium) at the same time.
With the description of the dynamics, gradual formation of core and corona in spatial and momentum space is achieved.
In the DCCI framework, the multiplicity dependence of the hadron yield ratios of multi-strange baryons to pions from small to large colliding systems in a wide range of collision energy is attributed to a continuous change of the fractions of the core and the corona components as multiplicity increases \cite{Kanakubo:2018vkl,Kanakubo:2019ogh}.
It should be noted here that the ``corona" is referred not only to an outer layer in the coordinate space but also to the one in the momentum space:
The lower $p_T$ partons are more likely to deposit their energy and momentum to form the fluids and the higher $p_T$ partons are less likely to be equilibrated during the DCCI processes.

In this paper, we update the DCCI framework towards a more comprehensive description of dynamics in full phase space from small to large colliding systems in a unified manner.
Hereafter we call this updated DCCI the \textit{DCCI2}.
In comparison with the previous work \cite{Kanakubo:2018vkl,Kanakubo:2019ogh}, several crucial updates have been made in this new version including a more sophisticated formula for four-momentum deposition of initial partons, particlization of the fluids on the switching hypersurface through a Monte-Carlo sampler \ISthreeD\  \cite{McNelis:2019auj}, hadronic rescatterings through a hadron cascade model \jam\  \cite{Nara:1999dz}, and modification of structure of color strings inside the fluids.
With these updates, the DCCI2 is capable of describing high-energy nuclear collisions from low to high $p_T$ region with particle identification in various colliding systems.

We generate initial partons from a general-purpose event generator \pythia8 \cite{Sjostrand:2007gs, Bierlich:2014xba}
switching off hadronization and make all of them sources of both the core and the corona parts. 
Here a special emphasis is put on to discriminate between the two terms, ``soft--hard" and ``core--corona". We call the core when it composes the matter in equilibrium. 
In DCCI2, the fluids generated through the dynamical initialization correspond to the core part and hardons particlized on switching hypersurface are regarded as the core components.
On the other hand, we call the corona when it composes matter completely out of equilibrium. 
In DCCI2, partons in the dilute regions and/or surviving even through the dynamical initialization correspond to the corona part, and hadrons from string decays are regarded as the corona components.
Although the core (corona) component is sometimes identified with the soft (hard) component, it is not the case in the DCCI2: The hadrons from string fragmentation are distributed all the way down to very low $p_T$ region, which one cannot consider as ``hard" components.
To the best of our knowledge, it has been believed so far without any strong justification that the corona components would be negligible in low $p_T$ region in heavy-ion collisions.
In this paper, we scrutinize 
the size of the contribution from the corona components 
in soft observables (``soft-from-corona") 
and how the fraction of the corona components evolves as multiplicity increases.

One of the main interests in this field is to constrain the transport coefficients of the QGP through the hydrodynamic analysis of anisotropic flow data in low $p_T$ regions.
It is conventionally assumed that the low $p_T$ hadrons are completely dominated by the core components. 
What if the corona components, whose contribution is often considered to be very small, affect the bulk observables in heavy-ion collisions?  
The non-equilibrium contribution is at most taken through small corrections to thermodynamic quantities such as shear stress and bulk pressure. 
Since the corona components are far-from-equilibrium distributions, these must be more important than those dissipative corrections in the hydrodynamic analysis if these are smaller than the core components, but of the same order of the magnitude.
Thus, dynamical modeling containing the core--corona picture could become the next-generation model inevitably needed for the precision study of the QGP properties. 

The present paper is organized as follow:
We explain details about the DCCI2 model in Sec.~\ref{sec:model}.
Staring from the general idea of dynamical initialization and the dynamical core--corona initialization,
we discuss some new features in the DCCI2 model.
In Sec.~\ref{sec:results}, we show the results from the DCCI2 model in p+p and Pb+Pb collisions at the LHC energies.
We also discuss the effects of string fragmentation and the longitudinal hydrodynamic expansion on the transverse energy per unit rapidity which play a crucial role in modeling hydrodynamic initial conditions. 
Section \ref{sec:summary} is devoted to the summary of the present paper.

Throughout this paper, 
we use the natural unit, $\hbar = c = k_{B} =1$, and the Minkowski metric,
$g_{\mu \nu} = \mathrm{diag}(1, -1, -1, -1)$.

\section{Model}
\label{sec:model}
The DCCI2 framework as a multi-stage dynamical model describes high-energy nuclear reactions from p+p to A+A collisions.
Before going into the details of the modeling of each stage, we briefly summarize the entire model flow of the DCCI2 framework.

\begin{figure}[htbp]
\begin{center}
\includegraphics[bb=0 0 576 1080, width=0.4\textwidth]{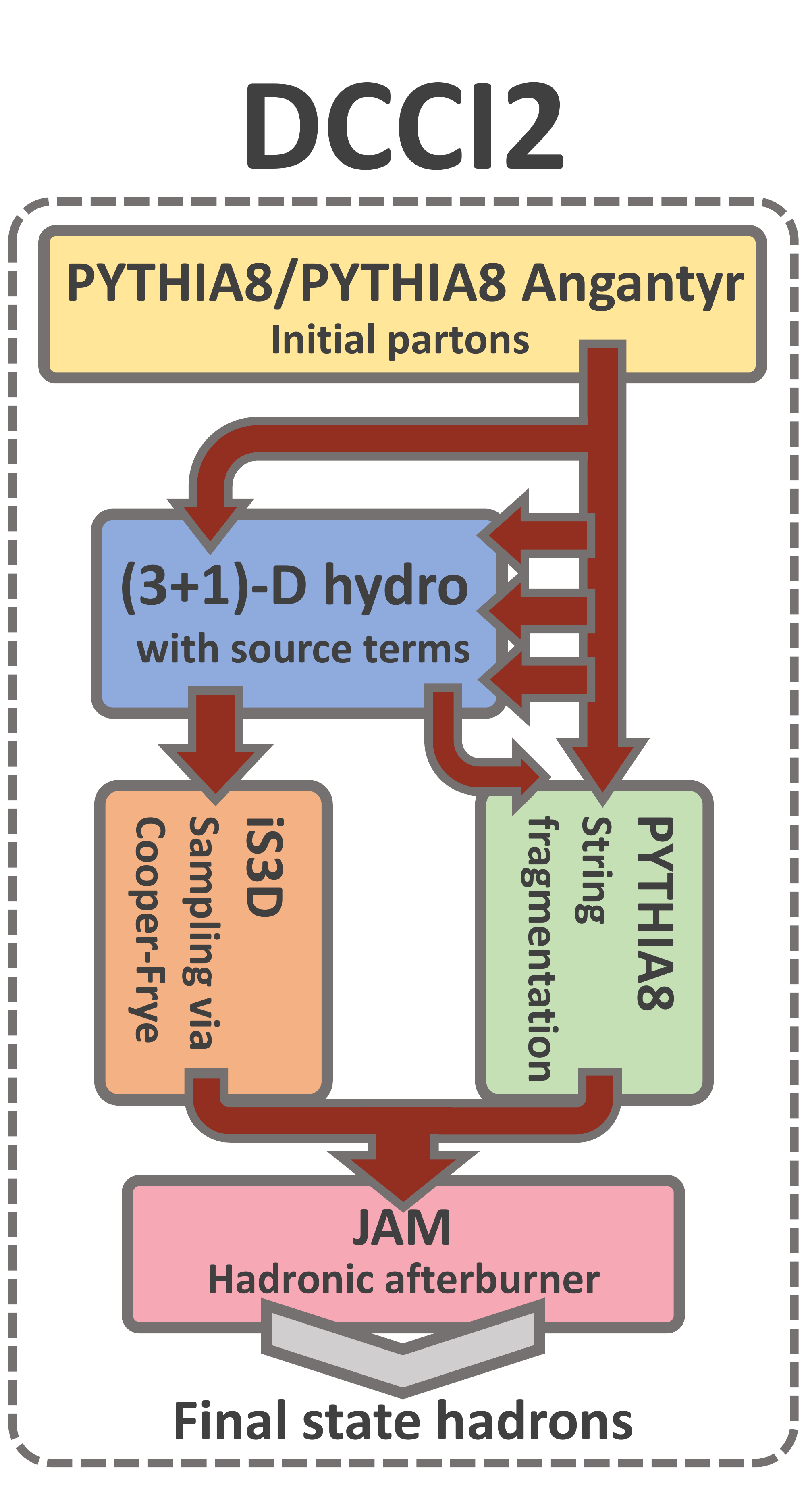}
\caption{(Color Online) Flowchart of the DCCI2 framework.
}
\label{fig:flow_chart}
\end{center}
\end{figure}

Figure~\ref{fig:flow_chart} represents the flowchart of the DCCI2 framework.
First, we obtain event-by-event phase-space distributions of initial partons produced just after the first contact of incoming nuclei using \pythia8.244 or its heavy-ion mode, Angantyr model \cite{Sjostrand:2007gs, Bierlich:2014xba}.
Hereafter, we call \pythia8 and \pythia8 Angantyr, respectively.
Those initial partons are assumed to be generated at a formation time, $\tau_{0}$.
Under the dynamical initialization framework,
the QGP fluids are generated via energy-momentum deposition from those initial partons by solving the relativistic hydrodynamic equations with source terms from $\tau = \tau_{0}$ to the end of the hydrodynamic evolution.
The energy-momentum deposition rate of partons is formulated based on the dynamical core--corona picture.
Partons which experience sufficient secondary interactions with surrounding partons are likely to deposit their energy-momentum and form QGP fluids. 
In contrast, partons that do not experience such sufficient secondary interactions give less contribution to the medium formation. 
Hydrodynamic simulations are performed in the (3+1)-dimensional Milne coordinates incorporating the $s95p$-v1.1 \cite{Huovinen:2009yb} equation of state (EoS). 
In the original $s95p$-v1, an EoS of the (2+1)-flavor lattice QCD at high temperature from HotQCD Collaboration \cite{Bazavov:2009zn} is smoothly connected to that from a hadron resonance gas, whose list is taken from Particle Data group as of 2004 \cite{Eidelman:2004wy}, at low temperature.
The particular version of EoS, $s95p$-v1.1, which we employ in the present calculations, is tuned to match the EoS of the hadron resonance gas with the resonances implemented in a hadronic cascade model, JAM, below a temperature of $184$~MeV.
The fluid elements below the switching temperature $T_{\mathrm{sw}}$
can be regarded as hadron gases whose evolution is described by the hadronic cascade model to be mentioned later.
Once the temperature of the fluid element goes down to $T(x) = T_{\mathrm{sw}}$, we switch the description from hydrodynamics to hadronic transport.
For the switch of the description, we use a Monte-Carlo sampler, \ISthreeD  \ \cite{McNelis:2019auj}, to convert hydrodynamic fields at the switching hypersurface to particles, which we call direct hadrons, in the EoS by sampling based on the Cooper-Frye formula \cite{Cooper:1974mv}.
Hadronization of non-equilibrated partons 
is performed by the string fragmentation in \pythia8. 
When a color string connecting partons from \pythia8 has a spatial overlap with the medium fluid, we assume that the string is cut and reconnected to 
partons sampled from the medium due to the screening effect of the medium.
The direct hadrons obtained from both \pythia8 and \ISthreeD \ are handed to the hadronic cascade model, \jam \ \cite{Nara:1999dz}, to perform hadronic rescatterings among them and resonance decays. 
In the following subsections, we explain the details of each stage.

\subsection{Generating initial partons}
\label{subsection:Generating_initial_partons}
The initially produced partons, \textit{i.e.}, all partons we use as an input of dynamical initialization, are obtained with \pythia8 or \pythia8 Angantyr.
Here we summarize settings that we use to generate initial partons.
\begin{itemize}
    \item {\texttt{PartonVertex:setVertex=on}}
    \item {\texttt{HadronLevel:all=off}}
    \item {\texttt{MultipartonInteractions:pT0Ref}}
    \item {\texttt{SpaceShower:pT0Ref}}
\end{itemize}
We basically use the default settings in \pythia8 and \pythia8 Angantyr to obtain phase-space distributions of partons except the two parameters, {\texttt{MultipartonInteractions:pT0Ref}}  and {\texttt{SpaceShower:pT0Ref}}.
These parameters regularize cross sections of multiparton interactions and infrared QCD emissions \cite{Pythia82OnlineManual}.
The same value of $p_{\mathrm{T0Ref}}$ is used for the both parameters just for simplicity.
Detailed discussion is given in Sec.~\ref{subsection:Evolution_of_transverse_energy}.

The information of color strings is given by \pythia8 and \pythia8 Angantyr besides phase-space information.
In order to respect the configuration of initially produced color strings, we keep this information for dynamical core--corona initialization.
Technically speaking, color and anti-color tags are given to each parton so that one is able to see the configuration of the color strings by tracing the tags.
Note that if there exist junctions, which are Y-shaped objects that three string pieces are converged, we keep this information as well to trace all strings generated in the event.

Eventually, we obtain a particle list for each event including particle IDs, phase-space information, and color and anti-color tags.
Junctions are added to the particle list if they exist in the event.
For heavy-ion collisions obtained with \pythia8 Angantyr, 
weighted events are generated \cite{Bierlich:2018xfw}: The impact parameter is distributed in a way that more central collisions and fewer peripheral collisions are generated than in the minimum bias cases. The corresponding weights are stored to be used in statistical analysis.

\subsection{Dynamical initialization}
\label{sebsec:DYNAMICAL_INITIALIZATION}
We phenomenologically and dynamically 
describe the initial stage of high-energy nuclear collisions through \textit{dynamical initialization}. 
Just after the first contact of incoming nuclei or nucleons, quarks and gluons are produced through hard scatterings or initial or final state radiations. 
Subsequently, some of those initially produced partons experience the secondary scatterings and contribute to forming the equilibrated matter.
On the other hand, partons that do not experience the interactions and partons surviving even after the secondary interactions contribute as the non-equilibrated matter.

To describe this stage, we start from the continuity equations of the entire system generated in a single collision event
\begin{align}
\label{eq:continuum_tot}
    \partial_\mu T^{\mu\nu}_{\mathrm{tot}} (x)=0.
\end{align}
If we assume that the entire system can be decomposed into equilibrated matter (fluids) and non-equilibrated matter (partons), Eq.~\refbra{eq:continuum_tot} can be written as 
\begin{align}
\label{eq:continuum_eq}
   \partial_\mu T^{\mu\nu}_{\mathrm{tot}}(x) =    \partial_\mu \left[ T^{\mu\nu}_{\mathrm{fluids}}(x) + T^{\mu\nu}_{\mathrm{partons}} (x)\right] = 0,
\end{align}
where $T^{\mu\nu}_{\mathrm{fluids}}$ and $ T^{\mu\nu}_{\mathrm{partons}}$ are energy-momentum tensors of equilibrated matter (fluids) and non-equilibrated matter (partons), respectively.
Then the space-time evolution of fluids can be expressed as a form of hydrodynamic equations with source terms
\begin{align}
\label{eq:hydro_with_source_terms}
\partial_\mu T^{\mu\nu}_{\mathrm{fluids}}(x) &= J^{\nu} (x),
\end{align}
where the source terms are written as
\begin{align}
\label{eq:source_terms}
J^\nu  = - \partial_\mu & T_{\mathrm{partons}}^{\mu\nu}.
\end{align}
Here we assume the energy and momentum deposited from non-equilibrated partons instantly reach a state under local thermal and chemical equilibrium. 

The exact form of the source terms is obtained by defining phase-space distributions and kinematics of initial partons \cite{Kanakubo:2019ogh}.
For the phase-space distributions, we assume
\begin{align}
\label{eq:phasespace-distribution-of-partons}
    f (\bm{x}, \bm{p}; t) d^3\!xd^3\!p = \sum_{i} G(\bm{x}\!-\!\bm{x}_i(t))  \delta^{(3)}\! (\bm{p}-\bm{p}_i(t))d^3\!xd^3\!p,
\end{align}
where $G(\bm{x}\!-\!\bm{x}_i(t))$ is a three-dimensional Gaussian distribution centered at a position of the $i$th parton, $\bm{x}_i(t)$, generated in one single event. 
We assume that each parton traverses along a straight trajectory. 
Under this assumption, the position of a parton at an arbitrary time is obtained as
\begin{align}
\label{eq:trajectories_of_partons}
    \bm{x}_{i} (t) = \frac{\bm{p}_{i}(t)}{p^0_{i}(t)}(t-t_{\mathrm{form},i})+\bm{x}_{\mathrm{form},i},
\end{align}
where 
$t_{\mathrm{form},i}$ and $\bm{x}_{\mathrm{form},i}$ are a formation time and a formation position, respectively. The $i$th parton is assumed to be formed at a common proper time, $\tau = \tau_{0}$. 
With these assumptions, the explicit form of the source terms in Eq.~\refbra{eq:source_terms} is obtained as \cite{Kanakubo:2019ogh}
\begin{align}
\label{eq:source_terms_dpdt}
    J^\nu &= - \partial_\mu T_{\mathrm{partons}}^{\mu\nu} \nonumber \\
    &=-\sum_{i}\int d^3 p \frac{p^\mu p^\nu}{p^0} \partial_\mu f(\bm{x},\bm{p}; t) \nonumber \\
    &=-\sum_{i}\frac{dp_i^\nu(t)}{dt}G(\bm{x}\!-\!\bm{x}_i(t)).
\end{align}
As one reads from the last line of Eq.~\refbra{eq:source_terms_dpdt}, 
the source terms of fluids are described as a summation of deposited energy-momentum of initial partons.

Space-time evolution of fluids
is described by ideal hydrodynamics.
This does not mean we do not describe any non-equilibrium components within DCCI2: Dissipative hydrodynamics deals only with small non-equilibrium corrections to equilibrium components, while the corona components, which are far from equilibrium, are taken into account in DCCI2. 
By neglecting dissipative terms for simplicity,
energy-momentum tensor of fluids is expressed as 
\begin{align}
    \label{eq:EMtensor_idealfluids}
    T_{\mathrm{fluids}}^{\mu\nu} = (e + P) u^{\mu}u^{\nu} - Pg^{\mu\nu},
\end{align}
where $e$, $P$, and $u^{\mu}$ are energy density, hydrostatic pressure, and four-velocity of fluids, respectively.
In this study, we do not solve conserved charges such as baryon number, strangeness, and electric charges.
It would be interesting to investigate them considering the initial distribution of the charges \cite{Martinez:2019jbu,Martinez:2019rlp}
.

As we mentioned at the beginning of this section, we perform actual hydrodynamic simulations in the (3+1)-dimensional Milne coordinates, $\tau = \sqrt{t^2-z^2}$, $\bm{x}_\perp = (x, y)$, and $\eta_s = (1/2) \ln{\left[(t+z)/(t-z)\right]}$.
In this case, the Gaussian distribution in Eq.~\refbra{eq:phasespace-distribution-of-partons} is replaced with
\begin{align}
 &G(\bm{x}_\perp\!-\!\bm{x}_{\perp, i}, \eta_{s}\!-\!\eta_{s,i}) d^2\!x_\perp \tau d\eta_{s}\nonumber \\ 
    & = \frac{1}{2\pi\sigma_{\perp}^2} \exp{\left[ -\frac{(\bm{x}_\perp - \bm{x}_{\perp,i} )^2}{2\sigma_\perp^2} \right]} \nonumber \\
& \times \frac{1}{\sqrt{2\pi\sigma_{\eta_s}^2 \tau^2}} \exp{\left[ -\frac{(\eta_{s} - \eta_{s,i})^2}{2\sigma_{\eta_s}^2} \right]}d^2\!x_\perp \tau d\eta_{s},
\end{align}
where $\sigma_{\perp}$ and $\sigma_{\eta_s}$ are transverse and longitudinal widths of the Gaussian distribution, respectively.
In the longitudinal direction, a straight trajectory implies $\eta_{s,i}=y_{p,i}$, where 
$\eta_{s,i} = (1/2) \ln{\left[(t+z_i)/(t-z_i)\right]}$ and $ y_{p,i}= (1/2)\ln{\left[(E_i+p_{z,i})/(E_i-p_{z,i})\right]}$ are space-time rapidity and momentum rapidity of the $i$th parton, respectively. 
In the following, we show some formulas in the Cartesian coordinates to avoid the complex representation of them. 
In any case, all the hydrodynamic simulations are performed in the Milne coordinates.

\subsection{Dynamical core--corona initialization}
\label{subsec:DYNAMICAL_CORECORONA_SEPARATION}
We establish the dynamical aspect of core--corona picture by
modeling the four-momentum deposition of partons $dp_i^\mu (t)/dt$ in Eq.~\refbra{eq:source_terms_dpdt} as an extension of the conventional core--corona picture.
The dynamical aspect of the core--corona picture, which we are going to model, is as follows:
Partons which are to experience sufficient secondary scatterings with others are likely to deposit their energy-momentum and form equilibrated matter (QGP fluids), while partons which are to rarely interact with others are likely to be free from depositing their energy-momentum.
To model the dynamical energy-momentum deposition under the core--corona picture, we invoke the equation of motion with a drag force caused by microscopic interactions with other particles.

We define the four-momentum deposition rate of the $i$th parton generated initially at a co-moving frame along $\eta_{s,i} = y_{p, i}$, space-time rapidity of the $i$th parton, as 
\begin{align}
\label{eq:four-momentum-deposition}
    \frac{d\tilde{p}_i^\mu}{d\tau} = - \sum_j^{\mathrm{coll}} \sigma_{ij}
     \tilde{\rho}_{ij}
    |\tilde{v}_{\mathrm{rel},ij}|\tilde{p}^\mu_i,
\end{align}
where $\sigma_{ij}$ is a cross section of the collision between the $i$th and $j$th partons, $ \tilde{\rho}_{ij}$ is an effective density of the $j$th parton seen from the $i$th parton which is normalized to be unity, $|\tilde{v}_{\mathrm{rel},ij}|$ is relative velocity between the $i$th and the $j$th partons. 
Variables with tilde are defined at each co-moving frame along $\eta_{s,i}$. 
The Lorentz transformation from laboratory frame to a co-moving frame along $\eta_{s,i}$ is given as, 
\begin{eqnarray} 
\label{eq:boost_eta} 
\Lambda^\mu_{\enskip \nu} (\eta_{s,i}) &=& \left( 
\begin{array}{cccc} 
\cosh{\eta_{s,i}} & 0 & 0 & -\sinh{\eta_{s,i}}\\ 
0 & 1 & 0 & 0\\ 
0 & 0 & 1 & 0\\ 
-\sinh{\eta_{s,i}} & 0 & 0 & \cosh{\eta_{s,i}} 
\end{array} 
\right). 
\end{eqnarray}

The summation in Eq.~\refbra{eq:four-momentum-deposition} is taken for all partons with non-zero energy that the $i$th parton will collide.
The candidate partons include not only initially produced ones but also ones in the thermalized medium.
We explain details on the treatment to pick up the thermalized partons in Sec.~\ref{subsection:Sampling_of_thermalized_partons}.
We employ an algorithm \cite{Hirano:2012yy} to evaluate the number of partonic scatterings that a parton undergoes along its trajectory.
Under the geometrical interpretation of cross sections, two partons, $i$ and $j$, are supposed to collide when the closest distance of them is smaller than $\sqrt{\sigma_{ij}/\pi}$ where $\sigma_{ij}$ is the same variable used in  Eq.~\refbra{eq:four-momentum-deposition}.
The cross section $\sigma_{ij}$ is defined as 
\begin{align}
\label{eq:parton_cross_section}
    \sigma_{ij}= \min\left\{ \frac{\sigma_0}{s_{ij}/\mathrm{GeV^2}}, \pi b_{\mathrm{cut}}^2 \right\},
\end{align}
where $\sigma_0$ is a parameter with a dimension of area, $s_{ij}$ is a Mandelstam variable $s_{ij} = (\tilde{p}_i^\mu + \tilde{p}_j^\mu)^2$, and $b_{\mathrm{cut}}$ is a parameter to avoid infrared divergence of the cross section when $s_{ij}$ becomes too small.
We neglect possible color Casimir factors in the cross section, and this parametrization is applied for all quarks, anti-quarks, and gluons.
It should be emphasized that this energy dependence of the cross section captures the core--corona picture in the \textit{momentum} space: the rare and the high-energy partons are not likely to deposit the four-momentum during the dynamical initialization process.

The effective density of the $j$th parton that is seen 
from the position of the $i$th one is defined as follows. The value of Gaussian
distribution centered at $\tilde{\bm{x}}_j$ is obtained at $\tilde{\bm{x}}_i$,
\begin{align}
 &    \tilde{\rho}_{ij}
  =\left. G(\tilde{\bm{x}}_{\perp}, \tilde{\eta}_{s}; \ \tilde{\bm{x}}_{\perp,j}, \tilde{\eta}_{s,j})\right|_{\tilde{\bm{x}}_{\perp}=\tilde{\bm{x}}_{\perp,i},  \tilde{\eta}_{s}= \tilde{\eta}_{s,i}}\nonumber \\
    & = \frac{1}{2\pi\tilde{\sigma}_{\perp}^2} \exp{\left[ -\frac{(\tilde{\bm{x}}_{\perp,i} - \tilde{\bm{x}}_{\perp,j} )^2}{2\tilde{\sigma}_\perp^2} \right]} \nonumber \\
    & \times \frac{1}{\sqrt{2\pi\tilde{\sigma}_{\eta_s}^2 \tau^2}} \exp{\left[ -\frac{(\tilde{\eta}_{s,i} - \tilde{\eta}_{s,j})^2
    }{2\tilde{\sigma}_{\eta_s}^2} \right]}.
\end{align}
Note that $\tilde{\eta}_{s,i} - \tilde{\eta}_{s,j} = \eta_{s,i} - \eta_{s,j}$, $\tilde{\bm{x}}_{\perp,i} = \bm{x}_{\perp,i}$, $\tilde{\sigma}_\perp = \sigma_{\perp}$, and $\tilde{\sigma}_{\eta_{s}} = \sigma_{\eta_{s}}$.
The relative velocity is calculated as
\begin{align}
    |\tilde{\bm{v}}_{\mathrm{rel}, \it{ij}}| = \left| \frac{\tilde{\bm{p}}_i}{\tilde{p}_i^0} - \frac{\tilde{\bm{p}}_j}{\tilde{p}_j^0} \right|.
\end{align}

As a consequence of the modeling for the four-momentum deposition rate, 
initial partons traversing dense regions with low energy and momentum tend to deposit their energy-momentum and generate QGP fluids.
On the other hand, initial partons traversing dilute regions with high energy tend to relatively keep their initial energy and momentum.
Here the factor $\sum_j^{\mathrm{coll}} \sigma_{ij}\tilde{\rho}_{ij} |\tilde{v}_{\mathrm{rel},ij}| d\tau$ can be regarded as the number of scattering that the $i$th parton experiences during $d\tau$.

During the DCCI processes, we monitor the change of the invariant mass of a string which is composed of the color-singlet combination of initial partons provided by \pythia8.
It should be noted that once the invariant mass of a string becomes smaller than a threshold to be hadronized via string fragmentation in \pythia8, energy-momentum of all the partons that compose the string is dumped into fluids.
In this model, we use $m_{\mathrm{th}} = m_1 + m_2 +1.0$ in units of $\mathrm{GeV}$ for the threshold, where $m_1 $ and $m_2$ are masses of each parton at both ends of the string. 

We emphasize here that the formulation of the four-momentum deposition rate of a parton is largely sophisticated from the one introduced in the previous work \cite{Kanakubo:2019ogh} although the basic concept of the core--corona picture is the same in both cases.
Under the previous work, there was a problem that high $p_T$ partons suffer from unexpected large suppression even in p+p collisions.
The reason is that, since partons in parton showers in a high $p_T$ jet are collimated and close to each other in both coordinate and momentum spaces, they had to deposit large four-momentum in our previous prescription in which only density and transverse momentum of the $i$th parton are taken into account \cite{Kanakubo:2019ogh}.
This problem is reconciled in this sophisticated modeling by considering trajectories of partons and relative velocities of parton pairs $|\tilde{v}_{\mathrm{rel},ij}|$.
Since trajectories of shower partons in a jet are supposed not to cross each other, the four-momentum deposition due to collisions among the shower partons is not likely to be counted in the summation in Eq.~\refbra{eq:four-momentum-deposition}.
Even if one consider trajectories of shower partons at the early time of dynamical core--corona initialization, they are close in space-time coordinates and would unreasonably deposit their four-momentum. 
The relative velocity avoids this issue because shower partons are supposed to have small relative velocities. 
Thus, because of the two factors, the dynamical core--corona initialization with Eq.~\refbra{eq:four-momentum-deposition} does not cause the unreasonable four-momentum deposition for partons in jets in DCCI2.

\subsection{Sampling of thermalized partons}
\label{subsection:Sampling_of_thermalized_partons}

As we mentioned in the previous subsection, the summation in the right hand side of  Eq.~\refbra{eq:four-momentum-deposition} is taken for all partons in the system including not only initially produced partons but also thermalized partons which are constituents of the QGP fluids.
This enables one to consider the four-momentum deposition due to scatterings with thermalized partons while traversing in the medium.
In order to consider the scatterings with thermalized partons, which are described by hydrodynamics, we sample partons from all fluid elements
and obtain phase-space distributions of them
at each time step.

Although we employ the lattice EoS, we make a massless ideal gas approximation on fluid elements for simplicity.
In this approximation, the number density of partons in a fluid element can be estimated as 
\begin{align}
    \label{eq:NumberDensity}
    n = \frac{90d^\prime \zeta(3)}{4\pi^4d}s_{\mathrm{EoS}}(T),
\end{align}
where $s_{\mathrm{EoS}}(T)$ is the entropy density obtained from the EoS via temperature $T$ at the fluid element.
The effective degeneracies of the QGP, $d$ and $d^\prime$, are defined as
\begin{align}
    \label{eq:dof_PartitionFunction}
    d &= d_{F} \times \frac{7}{8} + d_{B}, \\
    \label{eq:dof_NumberDensity}
    d^\prime & = d_{F} \times \frac{3}{4} + d_{B}.
\end{align}
The factors $\frac{7}{8}$ in Eq.~\refbra{eq:dof_PartitionFunction} and $\frac{3}{4}$ in Eq.~\refbra{eq:dof_NumberDensity} originate from differences between Fermi-Dirac and Bose-Einstein statistics in the entropy density and the number density.
The degrees of freedom of fermion $d_F$ and boson $d_B$ are obtains as
\begin{align}
    d_{F} &=d_{\rm{c}} \times d_{f} \times d_{s} \times d_{q\bar{q}} = 3\times3\times 2\times 2 = 36, \\
d_{B}&=d_{c} \times d_{s} = 8 \times 2 = 16,
\end{align}
where $d_{c}$, $d_{f}$, $d_{s}$, and $d_{q\bar{q}}$ represent the degrees of freedom of color, flavor, spin, and particle-antiparticle, respectively.

The number of partons in a fluid element is then, $\Delta N_0 = n \Delta x \Delta y  \tau \Delta \eta_s$,
where $n$ is the number density of partons obtained in Eq.~\refbra{eq:NumberDensity},
$\Delta x$, $\Delta y$, and $\Delta \eta_s$ are the widths of one fluid element in the Milne coordinates.
One can interpret $\Delta N_0$ as a mean value of Poisson distribution and sample the number of partons $N$ with
\begin{align}
 \dis P(N) = \exp{(-\Delta N_0)}\frac{\Delta N_0^{N}}{N!}.
\end{align}

For sampled $N$ partons, we stochastically assign species of them.
We pick up a quark or an anti-quark with a probability,
\begin{align}
 \label{eq:probability_of_samplingquark}
 P_{q/\bar{q}} &= \frac{(3/4) d_{F}}{(3/4)d_{F} + d_{B}},
\end{align}
which corresponds to a fraction of the degree of freedom of Fermi particles,
while a gluon is picked up with a probability, 
\begin{align}
 P_{g} &= 1- P_{q/\bar{q}}.
\end{align}

The three-dimensional momentum $\bm{k}$ of (anti-)quarks or gluons 
in the local rest frame of the fluid element is assigned according to the normalized massless Fermi or Bose distribution, 
\begin{align}
\label{eq:mass-lessBFdistribution_for_sampling}
    \dis P(\bm{k})d^3\!k = \frac{1}{N_{\mathrm{nom}}} \frac{1}{ \exp{(k / T)} \mp_{\mathrm{B, F}} 1}d^3\!k,
\end{align}
where $N_{\mathrm{nom}}$ is a normalization factor, $T$ is temperature of the fluid element, and $\mp_{\mathrm{B, F}}$ is a sign for Bose ($-$) and Fermi ($+$) statistics. Then, the energy and momentum in the lab frame is obtained by performing Lorentz transformation on $k^\mu=(\left|\bm{k}\right|,\bm{k})$ with the velocity of the fluid element.

Space coordinates are assigned with a uniform distribution within each fluid element.
For partons sampled from a fluid element centered at $\bm{x} = \bm{x}_i = \left(x_i, y_i, \eta_{s,i} \right)$, where the index $i$ stands for the numbering of fluid elements, we assign their coordinates with
\begin{align}
\label{eq:uniform_distribution}
& P_{\mathrm{uni}}(\bm{x})\tau \Delta x \Delta y \Delta \eta_{s} \nonumber \\
& = \left\{
    \begin{aligned}
         \  &0 \quad (\bm{x}<\bm{x}_i - \Delta \bm{x}/2, \ \bm{x}_i + \Delta \bm{x}/2  < \bm{x} )  \\
            &1 \quad (\bm{x}_i - \Delta \bm{x}/2 \leq \bm{x} \leq
            \ \bm{x}_i + \Delta \bm{x}/2 ) 
    \end{aligned}
    \right. .
\end{align}

As discussed in this subsection, the four-momentum deposition caused by collisions between a traversing parton as a corona part and thermalized partons as core parts could be regarded as a toy model of jet quenching.
We note that, although the implementation of a more sophisticated jet-quenching mechanism is the future work, the energy loss of traversing partons in the medium is phenomenologically introduced via the dynamical core--corona initialization in Eq.~\refbra{eq:four-momentum-deposition}.

\subsection{Modification of color strings}
\label{subsec:STRING_CUTTING}
The Lund string model is based on a linear confinement picture of color degree of freedom \cite{Andersson:1983ia,Sjostrand:2007gs}.
Energy stored between a quark and an anti-quark linearly increases with the separation length of the quark and anti-quark in the vacuum. 
However, this picture should be modified if we put them into the QGP at finite temperature.
It is known that, at high temperature, the string tension becomes so small that color strings would disappear \cite{Kaczmarek:1999mm}.
Since our input is initially produced partons connected with color strings and we generate fluids through their energy-momentum depositions, some color strings should overlap with the fluids in the coordinate space. 
We phenomenologically incorporate the modification of the color string configuration due to the finite temperature effect in DCCI2.

At $\tau = \tau_s(>\tau_{0})$, we assume that the string fragmentation happens when the entire color string is outside the fluids.
Here, ``a color string'' means chained partons as a color singlet object. 
When a color string is entirely inside the fluids at $\tau=\tau_s$, we discard the information of its color configuration and let its constituent partons evolve as individual non-equilibrated partons according to Eq.~\refbra{eq:trajectories_of_partons}. 
If a color string is partly inside and partly outside the fluids, the color string is subject to be cut off at the boundary of the fluids. The boundary here is identified with a contour of $T(x)=T_{\mathrm{sw}}$.
The color string in a vacuum cut off at the boundary is reattached to a thermal parton picked up from the hypersurface and forms a color-singlet object again to be hadronized via string fragmentation. The thermal parton is sampled in the hypersurface of the fluids. The details of the above treatment of color strings at $\tau = \tau_s$ are explained in Sec.~\ref{subsec:COLORSTRING_TREATMENT_TAU0}.
As for the rest of the color string left inside the fluids,
we discard the information of its color configuration and let its constituent partons evolve as individual non-equililbrated partons likewise the above case.

During the evolution of the fluids, the individual non-equilibrated partons come out from the hypersurface at some point. 
We assume that such a parton forms parton pairs to become a color-singlet object by picking up a thermalized parton from the hypersurface of the fluids. This prescription is based on exactly the concept of the coalescence models 
\cite{Hwa:2003ic,Hwa:2004ng,Greco:2003xt,Isse:2007pa,Fries:2008hs,Han:2016uhh,Fries:2019vws,Zhao:2019ehg,Kordell:2020wqi}. 
We explain the parton-pairing treatment at $\tau > \tau_s$ in Sec.~\ref{subsec:PARTON_PAIRING}.

We perform the modification only for color strings in which the transverse momentum of all partons forming that color string is less than a cut-off parameter, $p_T<p_{T, \mathrm{cut}}$, at $\tau=\tau_{0}$. Notice that this is merely a criterion of whether the modification of color string is performed and that all initially produced partons, including very high $p_T$ ones, nonetheless 
experience the dynamical core--corona initialization regardless of $p_{T,\mathrm{cut}}$.
This treatment avoids modification on $p_T$ spectra of final hadrons generated from intermediate to high $p_T$ partons which would less interact with fluids rather than low $p_T$ partons. 
Since the modification on the structure of color strings sensitively affect the final hadron distribution in momentum space, we should make a more quantitative discussion on the parameter $p_{T, \mathrm{cut}}$ as a future work.

It should also be noted that when we sample thermalized partons, energy-momentum is not subtracted from fluids just for simplicity.

\subsubsection{String cutting at $\tau=\tau_s$} 
\label{subsec:COLORSTRING_TREATMENT_TAU0}
We find a crossing point between a color string and the hypersurface of the fluids by tracing partons chained as color strings one by one. 
Since \pythia8 and \pythia8 Angantyr give us information of structure of color strings, we respect the initially produced color flow.

At $\tau = \tau_s$, all initial partons are classified into four types: ``hard" partons, dead partons inside fluids,  surviving partons inside fluids, and partons outside fluids.
We regard partons which is chained with at least one high $p_T (> p_{T,\mathrm{cut}})$ parton
as ``hard" partons.
We do not modify color strings composed of these hard partons to keep the initial color flow and hadronize them in a usual way discussed in Sec.~\ref{subsection:Direct_hadrons_from_core--corona_and_hadronic_afterburner}. 
During dynamical core--corona initialization, some partons
lose their initial energy completely inside fluids.
We regard those partons as dead ones and remove them from a list of partons.
These dead partons are no longer considered to be hadronized through string fragmentation.
For the other partons, we check whether they are inside the fluids one by one and regard them as surviving partons if it is the case.
These surviving partons are to be hadronized at $\tau>\tau_s$ if they have sufficient energy to come out from the fluids.
This be explained later in Sec.~\ref{subsec:PARTON_PAIRING}.
The rest of the partons are considered to be partons outside fluids.
Since partons outside the fluids cannot form color-singlet strings by themselves, we need to cut the original color strings at crossing points between the hypersurfaces and the color strings by sampling thermal partons.

In the following, we explain how to find crossing points and how to sample thermal partons.
We first assume that two adjacent partons in a color string, regardless of their status (dead, surviving, or outside fluids), are chained with linearly stretched color strings between the $i$th and the $(i+1)$th partons in coordinate space.
As a simple case, suppose that $[T(\bm{x}_{i})-T_{\mathrm{sw}}][T(\bm{x}_{i+1})-T_{\mathrm{sw}}] <0$,
where $\bm{x}_{i}= (x_{i}, y_{i}, \eta_{s,i})$ and $\bm{x}_{i+1}= (x_{i+1}, y_{i+1}, \eta_{s,i+1})$ are the positions of the $i$th and the $(i+1)$th partons, respectively, there exists 
a hypersurface of fluids between the $i$th and the ($i+1$)th partons.  
We scan temperature at all fluid elements along the linearly stretched color string from the $i$th to the $(i+1)$th parton to
find a crossing point.

Once the crossing point is found, a thermalized parton is picked up to form a color-singlet string.
The thermalized partons are sampled by using the information of the crossing point on the hypersurface such as velocity $\bm{v}_{\mathrm{hyp}}$, temperature $T_{\mathrm{hyp}}$, and coordinates $\bm{x}_{\mathrm{hyp}}$, which are obtained by taking an average of that of two adjacent fluid elements crossing the hypersurface. For instance, if the two adjacent fluid elements (symbolically denoted as the $j$th and the ($j+1$)th fluid element) have temperature $T_j>T_{\mathrm{sw}}$ and $T_{j+1}<T_{\mathrm{sw}}$, respectively, the temperature at the crossing point is obtained as $T_{\mathrm{hyp}} = (T_{j+1}+T_{j})/2$.
When the hypersurface of the fluids and the configuration of the color string are highly complicated, there could exist more than one crossing point between two adjacent partons. 
In such a case, we pick up a thermal parton from the closest crossing point for each parton in the string.

The species of the picked-up parton, whether if it is a quark, an anti-quark, or a gluon, is fixed by the configuration of color strings. 
For a color string that has a quark and an anti-quark at its ends, if string cutting removes the quark(anti-quark) side of the color string, an anti-quark(a quark) is picked up from the crossing point to form a color-singlet string in the vacuum.
When there is a color string that consists of two gluons while the only one of the gluons is inside of fluids, we pick up a gluon to make a color-singlet object.
On the other hand, for color strings with more than two gluons and no quarks or anti-quarks as their components, the so-called gluon loops, we cut the loop to open and pick up two gluons from the crossing points to make this a color-singlet object again.

A momentum of a picked-up parton is sampled with a normalized Fermi or Bose distribution,
\begin{align}
\label{eq:massive_bose_fermi}
    &P(\bm{p};m)d^3\!p  \nonumber \\ &=\frac{1}{N_{\mathrm{norm}}(m)} \frac{1}{\exp{\left[ \sqrt{\bm{p}^2 +m^2 } / T_{\mathrm{hyp}} \right] } \pm_{\mathrm{B,F}} 1}d^3\!p,
\end{align}
where 
\begin{align}
\label{eq:NormFactor}
   N_{\mathrm{norm}} (m)
    ={\dis\int \frac{1}{\exp{\left[ \sqrt{\bm{p}^2 +m^2 } / T_{\mathrm{hyp}} \right] } \pm_{\mathrm{B,F}} 1}d^3\!p}.
\end{align}
The energy of the (anti-)quarks is assigned so that they are mass-on-shell, which we require to perform hadronization via string fragmentation in \pythia8.
Four-momentum of these partons is Lorentz-boosted by using fluid velocity at the crossing point, $\bm{v}_{\mathrm{hyp}}$.

We stochastically assign flavors $f=u$, $d$,  or $s$ for each quark or anti-quark with the following probability,
\begin{align}
\label{eq:probability_flavor}
P_f & = N_{\mathrm{norm}} (m_f)/N_{\mathrm{sum}},\\
N_{\mathrm{sum}} & =N_{\mathrm{norm}}(m_u) +N_{\mathrm{norm}}(m_d)+N_{\mathrm{norm}}(m_s),
\end{align}
where the mass values of these quarks are taken from general settings in \pythia8.

As we mentioned in Sec.~\ref{subsec:DYNAMICAL_CORECORONA_SEPARATION}, there is a threshold of invariant mass of a color string to be hadronized via string fragmentation in \pythia8. If the invariant mass of a modified color string is smaller than the threshold, we remove partons forming the color string from a list of partons.

\subsubsection{Parton-pairing for surviving partons}
\label{subsec:PARTON_PAIRING} 
At $\tau>\tau_s$, we hadronize ``surviving partons traversing inside of the fluids'' when each of them comes out from fluids. To make the parton color-singlet to hadronize via string fragmentation, the parton picks up a thermal parton around the hypersurface.

Whether a parton comes out from medium or not is determined by the temperature of a fluid element at which the parton is currently located.
A surviving parton traverses a fluid according to Eq.~\refbra{eq:trajectories_of_partons}.

At the $k$th proper time step $\tau=\tau_k$, suppose that
a parton is at $\bm{x} = \bm{x}(\tau_k)$, where its temperature is $T(\bm{x}(\tau_k), \tau_k) > T_{\mathrm{sw}}$, and will move to $\bm{x} = \bm{x}(\tau_{k+1})$ at the next time step.

Simply assuming that the hypersurface does not change between the $k$th and the $(k+1)$th time step and see if the temperature satisfies $T(\bm{x}(\tau_{k+1}), \tau_k) < T_{\mathrm{sw}}$ by checking hypersurface only at the $k$th time step.
If the above condition is satisfied, the parton is regarded as coming out from the fluids at $\bm{x}_{\mathrm{hyp}} = [\bm{x}(\tau_k) + \bm{x}(\tau_{k+1})]/2$ at the $k$th time step.
For a quark (an anti-quark) coming out from medium, an anti-quark (a quark) is picked up to form a color-singlet string. On the other hand, for a gluon, a gluon is picked up to do so. A momentum is again sampled by using Eq.~\refbra{eq:massive_bose_fermi}, while its flavor is sampled with Eq.~\refbra{eq:probability_flavor}.

Note that, if a surviving parton fails to escape from the fluids by losing its initial energy completely, we regard that parton as a dead one  and remove it from a list of surviving partons.
Here again, if the invariant mass of the pair of partons is smaller than the threshold to be hadronized via string fragmentation in \pythia8, we remove them from the list of partons.

\subsection{Direct hadrons from core--corona and hadronic afterburner}
\label{subsection:Direct_hadrons_from_core--corona_and_hadronic_afterburner}
We switch the description 
of the hadrons in the core parts
from hydrodynamics to particle picture at the 
$T(x)=T_{\mathrm{sw}}$ hypersurface. 
The particlization of fluids is performed with \ISthreeD \ \cite{McNelis:2019auj}, which is an open-source code to perform conversion of hypersurface information of fluids into phase-space distributions of hadrons based on Monte-Carlo sampling of the Cooper-Frye formula \cite{Cooper:1974mv}.
Since the original \ISthreeD \ \cite{McNelis:2019auj} is not intended for event-by-event particlization,  
we extend the code so that this can be utilized for our event-by-event analysis.
We also change the list of hadrons in \ISthreeD \ to the one from the hadronic cascade model, \jam \ \cite{Nara:1999dz},  which is employed for the hadronic afterburner in the DCCI2 framework.
The hypersurface information is stored from $\tau = \tau_{0}$, the beginning of dynamical core--corona initialization, to the end of hydrodynamic evolution at which temperature of all fluid elements goes below $T_{\mathrm{sw}}$.
Fluid elements with $\bm{p}\cdot  d\bm{\sigma}<0$, which is known as the negative contribution in the Cooper-Frye formula, and those with $T<0.1 \ \mathrm{GeV}$ 
are ignored in \ISthreeD.
Note that ignoring the negative contributions makes it possible to count all flux generated via source terms in dynamical initialization.
In other words, if one integrates all flux including negative ones and neglects deposition of energy inside the fluids, total flux becomes zero due to Gauss's theorem
\footnote{In the conventional hydrodynamic models, initial conditions of hydrodynamic fields are put at a fixed initial time, $\tau = \tau_{\mathrm{init}}$, which can be regarded as a negative (in-coming) energy-momentum flux from the hypersurface  $T(\bm{x},\tau=\tau_{\mathrm{init}})=T_{\mathrm{sw}}$. Thus, thanks to the Gauss's theorem, the sum of out-going energy-momentum fluxes from the hypersurface $T(\bm{x}, \tau>\tau_{\mathrm{init}})=T_{\mathrm{sw}}$ is exactly the same as that of in-coming fluxes at $\tau = \tau_{\mathrm{init}}$ when there are no source terms in hydrodynamic equations.
}.
This is because, under the dynamical initialization framework, our simulation starts from the vacuum, and the deposited energy-momentum is regarded as incoming flux into the hypersurface.
We admit that energy-momentum conservation should be improved in a better treatment while we checked the conservation is satisfied within a reasonable range. 
The space-time coordinates of sampled hadrons $(x_i, y_i, \eta_{s,i})$ are assigned stochastically in the same way as one used for picking up thermalized partons explained in Sec.~\ref{subsection:Sampling_of_thermalized_partons}.

Regarding the corona parts, partons out of equilibrium undergo hadronization through string fragmentation.
Until $\tau=\tau_{s}$, we assume no hadronization occurs. At $\tau=\tau_{s}$, partons outside the fluids hadronize via string fragmentations. If a part of the string is inside the fluids, we modify the color string by cutting it at the crossing point as explained in Sec.~\ref{subsec:COLORSTRING_TREATMENT_TAU0} and hadronize the modified color string.
Once surviving partons come out from the hypersurface of fluids after $\tau=\tau_{s}$, 
those partons are hadronized by picking up a thermal parton to form a string as discussed in \ref{subsec:PARTON_PAIRING}.
The string composed of at least one high $p_{T} (> p_{T,\mathrm{cut}})$ parton is hadronized when all the partons chained with this high $p_{T}$ parton come out from the fluids. 

The string fragmentation is performed by utilizing  \pythia8.
The flag \texttt{ProcessLevel:all=off} is set to stop generating events and \texttt{forceHadronLevel()} is called to perform hadronization against the partons manually added as an input.
The information of input partons handed to \pythia8 is,
particle ID, four-momentum, coordinates, color, and anti-color. 
As for coordinates, only transverse coordinates, $x$ and $y$, of partons are handed while $t$ and $z$ are set to be zeros.
This is because assigning $t$ and $z$ may cause violation of causality and should be treated carefully in \pythia8.
We correct the energy of the partons to be mass-on-shell using their momenta and rest masses to perform string fragmentation in \pythia8. This is the same procedure as we did in the previous work \cite{Kanakubo:2019ogh} since quarks or anti-quarks that lose their energy in dynamical initialization are mass-off-shell due to the four-momentum deposition of Eq.~\refbra{eq:four-momentum-deposition}. 
Information of vertices of generated hadrons are obtained with an option \texttt{Fragmentation:setVertices = on} based on the model proposed in Ref.~\cite{Ferreres-Sole:2018vgo}. We use this information for initial conditions in \jam.

In both the particlization by \ISthreeD \ 
and the string fragmentation by \pythia8, 
we turn off decays of unstable hadrons. 
Instead, JAM handles decays of the unstable hadrons together with rescatterings while describing their space-time evolution in the late stage. 
The hadrons obtained from both \ISthreeD \ and \pythia8 are put into \jam \ all together to perform the hadronic cascade since both components should interact with each other.
In \jam, an option to switch on or off hadronic rescatterings is used to see its effect on final hadrons. It should be also noted that we turn off electroweak decays except $\Sigma^0 \rightarrow \Lambda + \gamma$ to directly compare our results with experimental observable.

\subsection{Parameter set in DCCI2}
\label{subsection:Parameter_set_in_DCCI2}
Here we summarize all the parameters that we use throughout this paper.
\begin{table}[htpb]
\caption{Parameter set in DCCI2 used throughout this paper.
}
\begin{center}
\begin{tabular}{cc} 
\hline \hline 
 \hspace{0.5cm} Parameters \hspace{0.5cm} & \hspace{0.5cm} values  \hspace{0.5cm} \\
\hline
$p_{\mathrm{T0Ref}}$ (p+p) & 1.8 GeV \\
$p_{\mathrm{T0Ref}}$ (Pb+Pb) & 0.9 GeV \\
$\tau_{0}$ & 0.1 fm \\
$\tau_{s}$  & 0.3 fm \\
$T_{\mathrm{sw}}$  & 0.165 GeV \\
$\sigma_0$ & 0.4 $\mathrm{fm}^2$ \\
$b_{\mathrm{cut}}$ & 1.0 fm \\
$p_{T,\mathrm{cut}}$ & 3.0 GeV \\
$\sigma_\perp$  & 0.5 fm \\
$\sigma_{\eta_s}$  & 0.5 \\
$\Delta x$ & 0.3 fm \\
$\Delta y$ & 0.3 fm \\
$\Delta \eta_s $ & 0.15 \\ 
\hline \hline
\end{tabular}
\end{center}
\label{tab:PARAMETERSET}
\end{table}

Note that we use the same parameters for both p+p and Pb+Pb collisions except $p_{\mathrm{T0Ref}}$.
In the conventional hydrodynamic models, several parameters are used to directly parametrize the initial profiles of hydrodynamic fields. 
In contrast in DCCI2, how many initial partons are generated is determined in 
\pythia8 or \pythia8 Angantyr and how much the energy-momentum of these initial partons are converted to the hydrodynamic fields is controlled through the parameters, $\sigma_0$, $b_{\mathrm{cut}}$, $\sigma_\perp$, and $\sigma_{\eta_s}$, in Eq.~\refbra{eq:four-momentum-deposition}.
More details on how to fix these parameters are discussed in Sec.~\ref{sec:results}.

\section{Results and Discussions}
\label{sec:results}
In this paper, we simulate p+p collisions at $\sqrt{s}  = 7$ TeV and Pb+Pb collisions at $\sqrt{s_{NN}} = 2.76$ TeV with DCCI2. 
The following results are obtained from full simulations of 300K and 12.5K events for p+p and Pb+Pb collisions, respectively.
In Sec.~\ref{subsection:PARAMETER_DETERMINATION},
we start with fixing some major parameters in DCCI2 to reproduce the experimental data of the charged particle multiplicity as functions of multiplicity (p+p) or centrality (Pb+Pb) classes and
the multiplicity dependence in particle yield ratios of omega baryons to charged pions.
As a result of the parameter determination, fractions of core and corona components to final hadronic productions as a function of charged particle multiplicity at midrapidity are extracted. 
Next, we show the transverse momentum spectrum in p+p and Pb+Pb collisions and its breakdown into core and corona components in Sec.~\ref{sec:TRANSVERSE_MOMENTUM_CORECORONA}.
In order to see the interplay between core and corona components on observable obtained from final hadrons, we analyze the mean transverse momentum and the second-order anisotropic flow coefficients 
as functions of the number of produced charged particles in certain kinematic windows in Sec.~\ref{sec:CORECORONA_CONTRIBUTION}.
Finally, we show multiplicity dependence of radial flow effects based on violation of the mean transverse mass scaling
and discuss if the effect can be discriminated from the one originating from pure string fragmentation with color reconnection \cite{Sjostrand:2004pf} in Sec.~\ref{sec:MTSCALING}.
Due to the two competing particle production mechanisms,
it is not trivial to reproduce the multiplicity within a 
two-component model like DCCI2. We discuss details of this issue in Sec.~\ref{subsection:Evolution_of_transverse_energy}.

Let us note that the effects of string cutting explained in Sec.~\ref{subsec:STRING_CUTTING} are not investigated throughout this paper because the modification on color strings adds another complexity which we want to avoid in this discussion.

\begin{figure*}[htbp]
 \begin{center}
\includegraphics[bb=0 0 967 1047, width=1.0\textwidth]{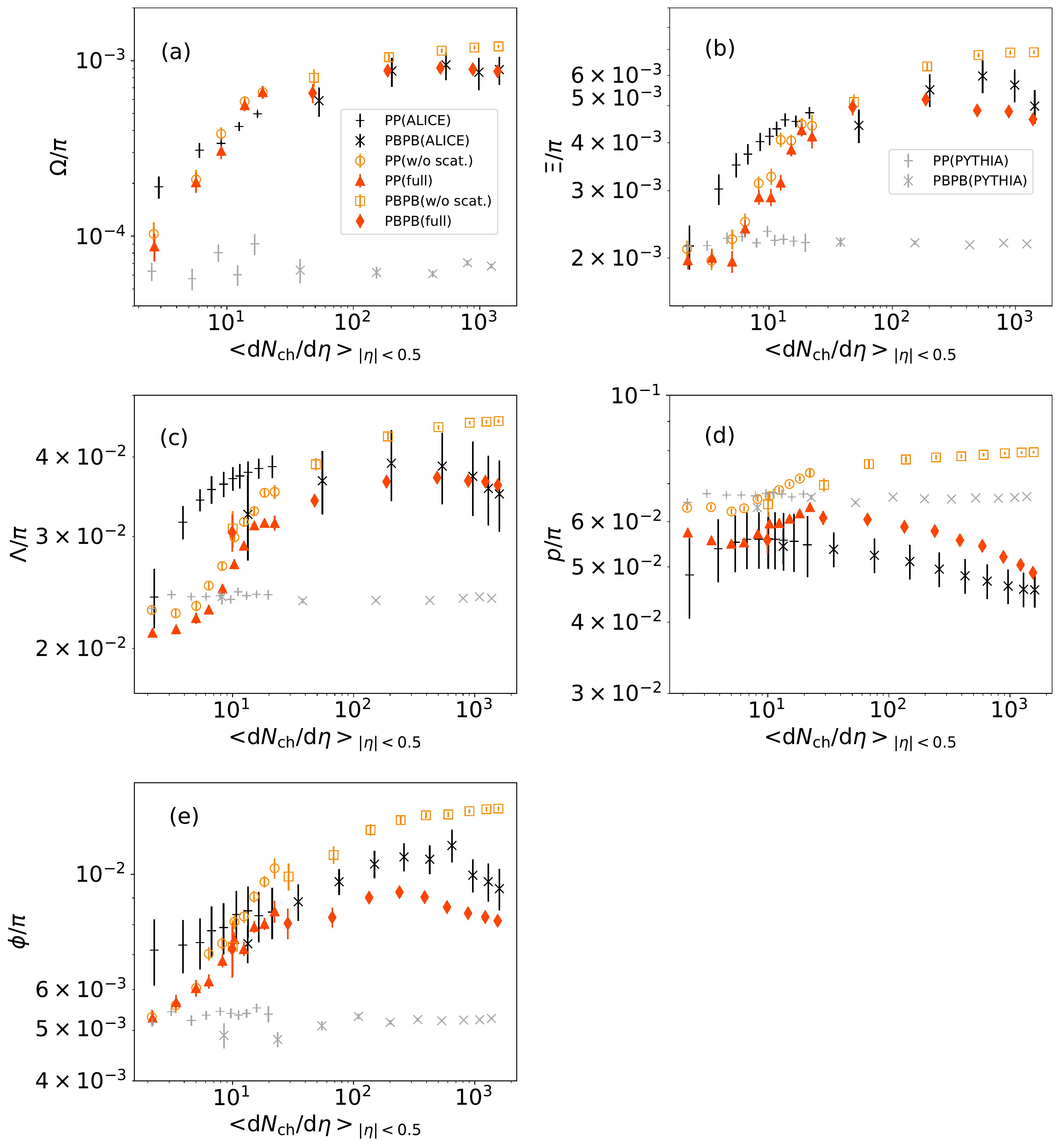}
  \caption{(Color Online)
  Particle yield ratios of (a) omegas ($\Omega^-$ and $\bar{\Omega}^{+}$), (b) cascades ($\Xi^{-}$ and $\bar{\Xi}^{+}$), (c) lambdas ($\Lambda$ and $\bar{\Lambda}$), (d) protons ($p$ and $\bar{p}$), (e) phi mesons ($\phi$) to charged pions ($\pi^+$ and $\pi^-$) as functions of charged particle multiplicity at midrapidity in p+p and Pb+Pb collisions.
  Results from full simulations of DCCI2 in p+p collisions at $\sqrt{s}=7$ TeV (closed red triangles) and Pb+Pb collisions at $\sqrt{s_{NN}} = 2.76$ TeV (closed red diamonds) collisions are compared with the ALICE experimental data in p+p (black pluses) and Pb+Pb (black crosses) collisions
  \cite{ALICE:2017jyt, ABELEV:2013zaa, Abelev:2013haa, Acharya:2018orn, Abelev:2013vea, Abelev:2014uua}.  
  The $\Lambda/\pi$ ratio in Pb+Pb collisions at $\sqrt{s_{NN}}=2.76 \ \mathrm{TeV}$
   reported by the ALICE Collaboration in Ref.~\cite{Abelev:2014uua}
  is plotted as a function of the number of participants $N_{\mathrm{part}}$ rather than charged particle multiplicity.
  The corresponding charged particle multiplicity at midrapidity is taken from Ref.~\cite{Abelev:2013vea}.
  Results without hadronic rescatterings are also plotted in p+p (open orange circles) and Pb+Pb (open orange squared)  collisions.
  Results from \pythia8 in p+p collisions (gray pluses) and from \pythia8 Angantyr in Pb+Pb (gray crosses) collisions are plotted as references.}
  \label{fig:PARTICLRRATIO_PP_PBPB}
 \end{center}
\end{figure*}

\subsection{Parameter determination and fractions of core and corona components}
\label{subsection:PARAMETER_DETERMINATION}

Here we focus on two main parameters in DCCI2, 
$p_{\mathrm{T0Ref}}$ used in the generation of initial partons in \pythia8 and \pythia8 Angantyr, and $\sigma_0$ to scale the magnitude of cross sections in Eq.~\refbra{eq:four-momentum-deposition}.
We determine these parameters to reasonably describe both 
the charged particle multiplicity as a function of multiplicity (p+p) or centrality (Pb+Pb) classes at midrapidity
and particle yield ratios of omega baryons to charged pions as functions of charged particle multiplicity.

The multi-strange hadron yield ratios tell us fractions of contributions from thermalized (core) and non-thermalized (corona) matter to total final hadron yields \cite{Kanakubo:2018vkl,Kanakubo:2019ogh}. 
On the other hand, the charged particle yields need to be used in the parameter determination together with the particle yield ratios. 
These two parameters are highly sensitive to both charged particle multiplicity and particle yield ratios and are strongly correlated. 
Detailed discussion on this issue is made in Sec.~\ref{subsection:Evolution_of_transverse_energy}. 
The resultant parameter values are summarized in Table \ref{tab:PARAMETERSET} in Sec.~\ref{subsection:Parameter_set_in_DCCI2}.
Here, the switching temperature $T_{\mathrm{sw}}$, which controls particle yield ratios as the parameters mentioned above do, is fixed to describe the ratios of omega baryons to charged pions in central Pb+Pb collisions. 

\begin{figure*}[htbp]
  \begin{center}
 \includegraphics[bb=0 0 552 537, width=0.45\textwidth]{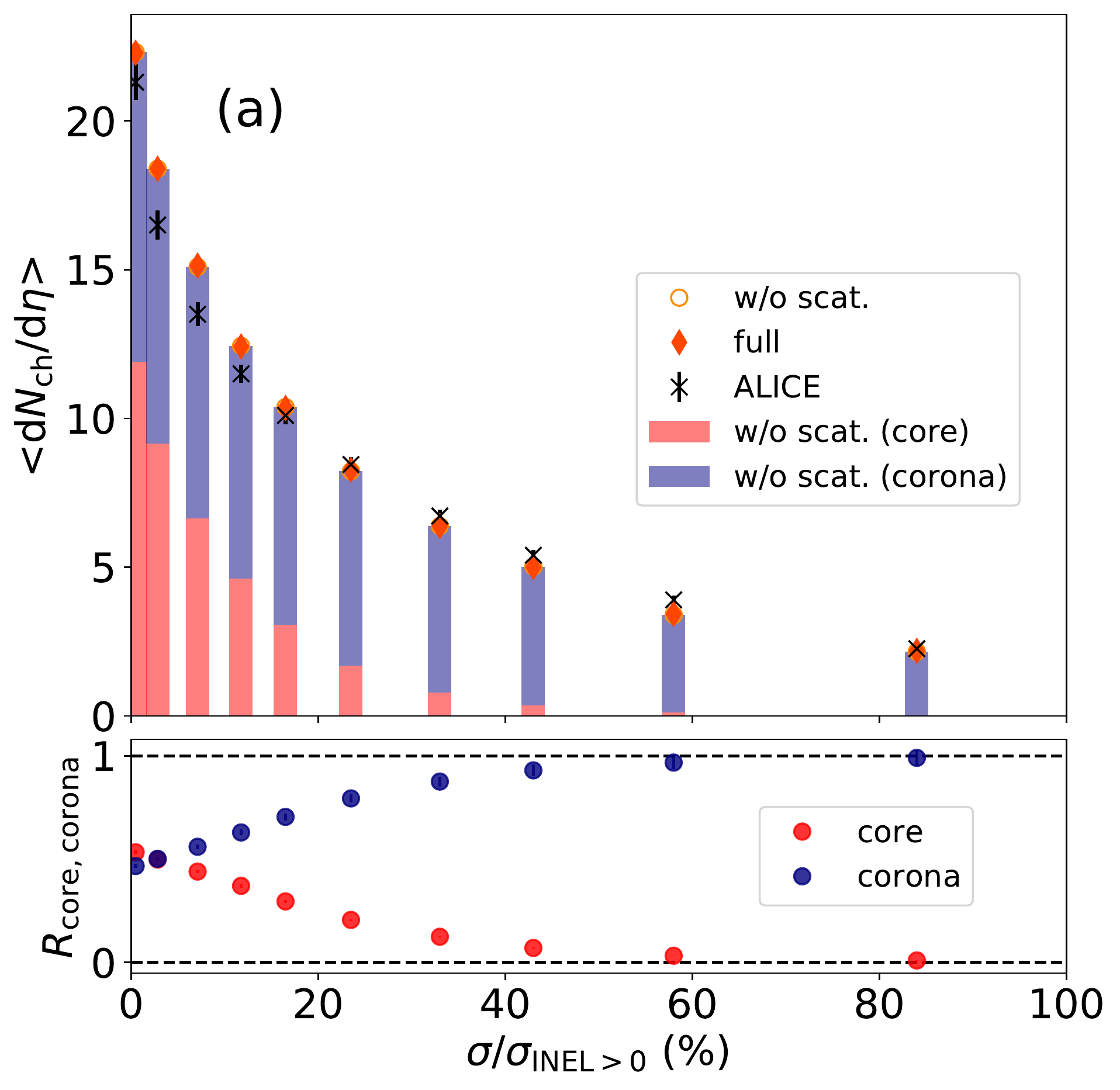}
\includegraphics[bb=0 0 552 537, width=0.45\textwidth]{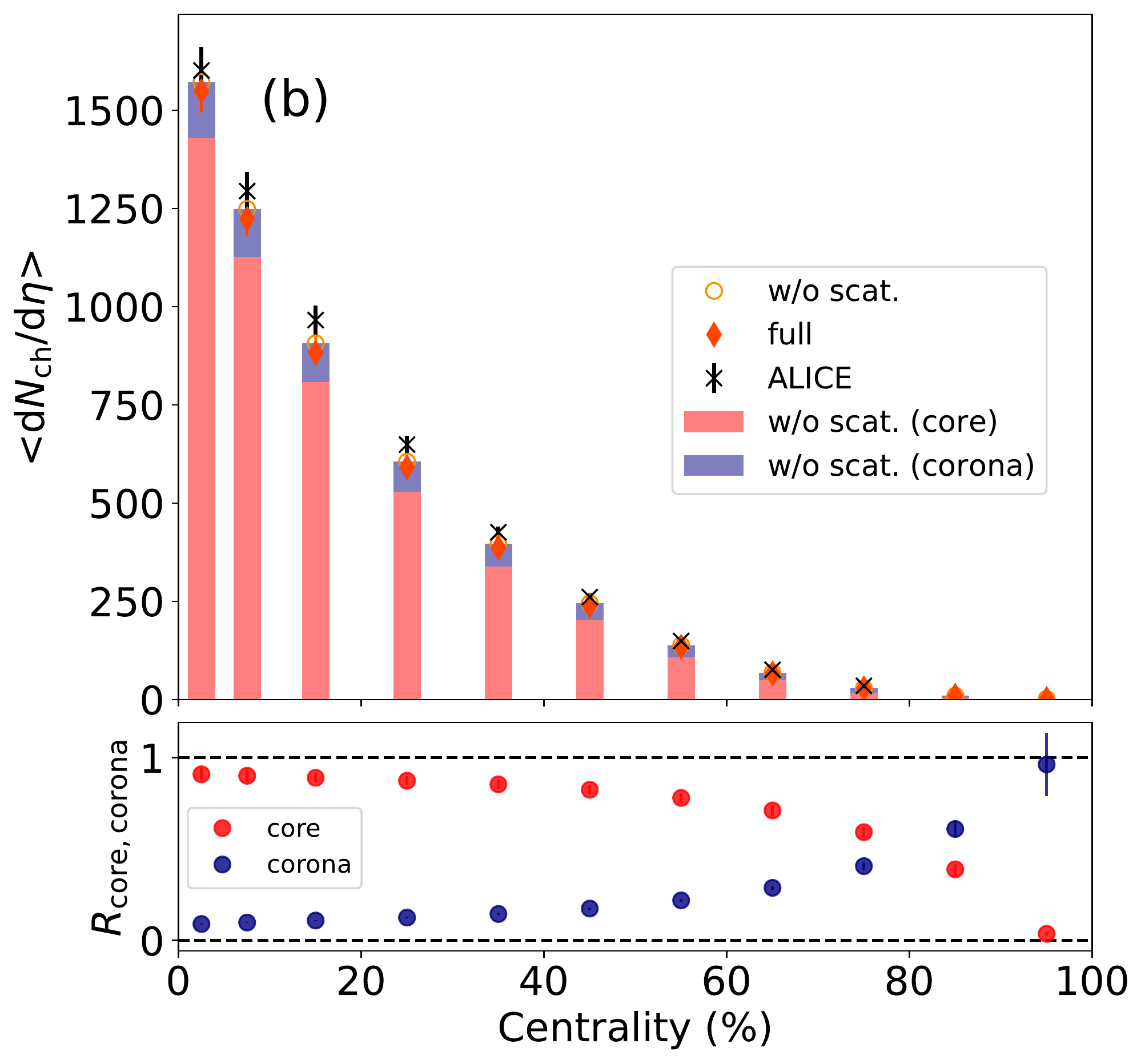}
\caption{(Color Online)
(Upper) (a) Charged particle multiplicity at midrapidity as a function of the fraction of the INEL$>0$ cross sections in p+p collisions at $\sqrt{s} = 7$ TeV.
(b) Charged particle multiplicity at midrapidity as a function of centrality from Pb+Pb collisions at $\sqrt{s_{NN}} = 2.76 \ \mathrm{TeV}$.
Results from full simulations (orange diamonds) and simulations without hadronic rescatterings (open orange circles) are compared with the ALICE experimental data (black crosses) \cite{Acharya:2018orn,Aamodt:2010cz}.
Core and corona contributions without hadronic rescatterings are shown in red and blue stacked bars, respectively.
(Lower) Fractions of core (red) and corona (blue) components 
without hadronic rescatterings are plotted in (a) p+p and (b) Pb+Pb collisions.
}
\label{fig:MULTIPLICITY_PP_PBPB}
 \end{center}
\end{figure*}

Figure~\ref{fig:PARTICLRRATIO_PP_PBPB} shows particle yield ratios to charged pions produced in $|y|<0.5$ as functions of charged particle multiplicity $|\eta|<0.5$ in p+p and Pb+Pb collisions compared with the ALICE experimental data \cite{ALICE:2017jyt, ABELEV:2013zaa, Abelev:2013haa, Acharya:2018orn, Abelev:2013vea, Abelev:2014uua}.
Note that the charged particle multiplicity at midrapidity $\langle dN_{\mathrm{ch}}/d\eta \rangle_{|\eta|<0.5}$ in the horizontal axes is obtained by using V0M ($-3.7< \eta < -1.7$ and $2.8 <\eta < 5.1$) multiplicity (p+p) or centrality (Pb+Pb) class, which is the same procedure as used in the ALICE data \cite{Abelev:2014ffa}.
Determining these classes by using the multiplicity in forward and backward rapidity regions is essential even in theoretical analysis to avoid the effect of self-correlation on observables at midrapidity \cite{Acharya:2018orn}.
Throughout this paper, the ``charged particles'' mean the sum of charged pions, charged kaons, protons, and antiprotons, which do not contain contributions from weak decays.
To obtain particle ratios of primary strange hadrons, which are stable against strong decays, we switch off their weak decays in JAM. 
Note that we take into account a particular electromagnetic decay, $\Sigma^0 \rightarrow \Lambda + \gamma$, in the presented results of $\Lambda$ yields \cite{ALICEPrimaryParticle}.
Results with switching off hadronic rescatterings are shown to reveal the effect of hadronic rescatterings \cite{Hirano:2005xf,Hirano:2007ei,Takeuchi:2015ana} on both core and corona components in the late stage.
Results from \pythia8 for p+p collisions and \pythia8 Angantyr for Pb+Pb are also plotted as references.

Overall, smooth changes of the particle yield ratios are observed along charged particle multiplicity, which is consistent with our previous studies \cite{Kanakubo:2018vkl,Kanakubo:2019ogh}. 
Due to the implementation of the core--corona picture in the dynamical initialization framework, particle productions from corona components with string fragmentation are dominant in final hadron yields in low-multiplicity events, while those from core components produced from equilibrated matter are dominant in high-multiplicity events.
Thus the overall tendency is that the particle yield ratio at low-multiplicity events almost reflects its value obtained from string fragmentation, while the one at high-multiplicity events reflects the value obtained only from hadronic productions from hydrodynamics.
Notice that the particle yield ratios of all hadronic species are almost independent of multiplicity from p+p to Pb+Pb collisions with default \pythia8 and \pythia8 Angantyr respectively, which is one of the manifestations of  ``jet universality", namely, 
the string fragmentation being independent of how the string is formed from e$^+$+e$^{-}$ to Pb+Pb collisions \cite{Buckley:2011ms} 
\footnote{Note that \pythia 8 and \pythia 8 Angantyr with rope hadronization show enhancement of strange hadron yield ratios as a function of multiplicity
\cite{Bierlich:2014xba,Bierlich:2017sxk}.}.

We tune the parameters in the full simulations of DCCI2 to reasonably reproduce
the particle yield ratios of omega baryons to pions, $\Omega/\pi$,
reported by the ALICE Collaboration \cite{ALICE:2017jyt,ABELEV:2013zaa}.
Although we have to admit that our results do not perfectly describe the experimental data as one sees in Fig.~\ref{fig:PARTICLRRATIO_PP_PBPB} (a), fine-tuning of the parameters is beyond the scope in this paper.
In Figs.~\ref{fig:PARTICLRRATIO_PP_PBPB} (b)-(e),
we also show results of cascades, lambdas, protons, and phi mesons, respectively. 
For the ratios of cascades to pions, $\Xi/\pi$, in Fig.~\ref{fig:PARTICLRRATIO_PP_PBPB} (b), our results underestimate the experimental data except the lowest- and the highest-multiplicity classes in p+p collisions.
For the ratios of lambdas to pions, $\Lambda/\pi$,  in Fig.~\ref{fig:PARTICLRRATIO_PP_PBPB} (c), our results from full simulations show smaller values than the experimental data in p+p collisions for almost the entire charged particle multiplicity, while it shows good agreement with the data in Pb+Pb collisions. For the ratios of protons to pions, $p/\pi$, in Fig.~\ref{fig:PARTICLRRATIO_PP_PBPB} (d), 
our full results including hadronic rescatterings through JAM 
qualitatively describe the decreasing behavior along the charged particle multiplicity in the experimental data in Pb+Pb collisions. 
This is consistent with a perspective of proton-antiproton annihilations \cite{Werner:2018yad,Abelev:2013vea}. 
The annihilation effect is seen even in p+p collisions, which leads to a better agreement with the experimental data. 
For the ratios of phi mesons to pions, $\phi/\pi$,  in 
Fig.~\ref{fig:PARTICLRRATIO_PP_PBPB} (e), the tendency in the experimental data above $\langle dN_{\mathrm{ch}}/d\eta \rangle_{|\eta|<0.5} \sim 7$ is well captured by our full result. 
In particular, the dissociation of phi mesons in hadronic rescatterings plays an important role to describe the suppression at high-multiplicity as observed in the experimental data.

Notably, the increasing behavior along charged particle multiplicity in p+p collisions is achieved in our results with the core--corona picture.
It is also discussed that
canonical suppression models, which are commonly used in the discussion on multiplicity dependence of particle yield ratios in comparison with the core--corona picture,
need to incorporate incompleteness of chemical equilibrium for strangeness due to the hidden strangeness of phi mesons \cite{Sollfrank:1997bq, Vovchenko:2019kes}.
Thus the increasing behavior would clearly represent that the matter formed in p+p collisions is under incompleteness of chemical equilibrium for strangeness.

The upper panel of Fig.~\ref{fig:MULTIPLICITY_PP_PBPB} (a) shows charged particle multiplicity at midrapidity $\langle  dN_{\mathrm{ch}}/d\eta \rangle_{|\eta|<0.5}$ as a function of multiplicity class $\sigma/\sigma_{\mathrm{INEL}>0}$ 
in p+p events. Here, we take into account only $\mathrm{INEL} >0$
events in which at least one charged particle is produced within a pseudorapidity range $|\eta|<1.0$ defined in the ALICE experimental analysis \cite{Acharya:2018orn}.

The upper panel of Fig.~\ref{fig:MULTIPLICITY_PP_PBPB} (b) shows the same observable but as a function of centrality class in Pb+Pb collisions.
Here we again note that each multiplicity or centrality class is obtained with V0M multiplicity.
In both figures, results from simulations with and without hadronic rescatterings are compared with the ALICE experimental data \cite{Acharya:2018orn,Aamodt:2010cz}.
Each contribution from core and corona components is separately shown as stacked bars for the case without hadronic rescatterings.
It should be noted that the separation of core and corona components in DCCI2 is attained only by switching off hadronic rescatterings in JAM.
This is because hadronic rescatterings mix those two components up by causing parton exchange between hadrons or formation of excited states. 
Both of our results show the reasonable description of the ALICE experimental data in p+p and Pb+Pb collisions.
From the comparison between with and without hadronic rescatterings, (quasi-)elastic scatterings would be dominant and, as a result, the effect of hadronic rescatterings on multiplicity turns out not to be significant.

The lower panels of Figs.~\ref{fig:MULTIPLICITY_PP_PBPB} (a) and (b) show the yield fractions of core and corona components to the total from results without hadronic rescatterings, $R_{\mathrm{core}}$ and $R_{\mathrm{corona}}$, respectively, as functions of multiplicity (p+p) and centrality (Pb+Pb) classes.
Smooth changes along multiplicity and centrality classes are observed in both p+p and Pb+Pb collisions. 
In Fig.~\ref{fig:MULTIPLICITY_PP_PBPB} (a),
the fraction of core components in p+p collisions almost vanishes for 48-68\% and 68-100\% multiplicity classes, in which $\langle dN_{\mathrm{ch}}/d\eta \rangle_{|\eta|<0.5}$ is less than $\sim5$.
Then, it increases along multiplicity and reaches $R_{\mathrm{core}} \sim 0.53$ in the highest multiplicity class 0.0-0.95\% in which $\langle dN_{\mathrm{ch}}/d\eta \rangle_{|\eta|<0.5} \sim 21$. 
One also sees that the contribution of core components overtakes that of corona components only in 0.0-0.95\%  multiplicity class
within our calculations with the current parameter set.
This supports a perspective that recent observations of collectivity in high-multiplicity small colliding systems at the LHC energies result from the (partial) formation of the QGP fluids.
It should also be noted that the fraction of core components shows $R_{\mathrm{core}}\sim0.12$ at $\langle dN_{\mathrm{ch}}/d\eta \rangle_{|\eta|<0.5} \sim 7$ which is minimum-bias multiplicity for $\mathrm{INEL} >0$ events \cite{Adam:2015gka} \footnote{
The result from EPOS 3.210 shows $\sim 30\%$ at the same multiplicity \cite{Werner:2019aqa}.}.
The lower panel of Fig.~\ref{fig:MULTIPLICITY_PP_PBPB} (b) shows results in Pb+Pb collisions.
The core components highly dominate, $R_{\mathrm{core}} \gtrsim 0.90$, from 0 to 10\% centrality classes where their corresponding multiplicities are above $\langle dN_{\mathrm{ch}}/d\eta \rangle_{|\eta|<0.5} \sim 10^3$. 
The corona components become dominant around at $80$\% centrality class towards peripheral events.
It should also be mentioned that the contribution of corona components remains $R_{\mathrm{corona}}\sim 0.17$-$0.22$ at midrapidity in intermediate centrality classes ($\sim40$-$60\%$) where the whole systems is often assumed to be described by hydrodynamics.

\begin{figure}[tbp]
\begin{center}
\includegraphics[bb=0 0 645 447, width=0.5\textwidth]{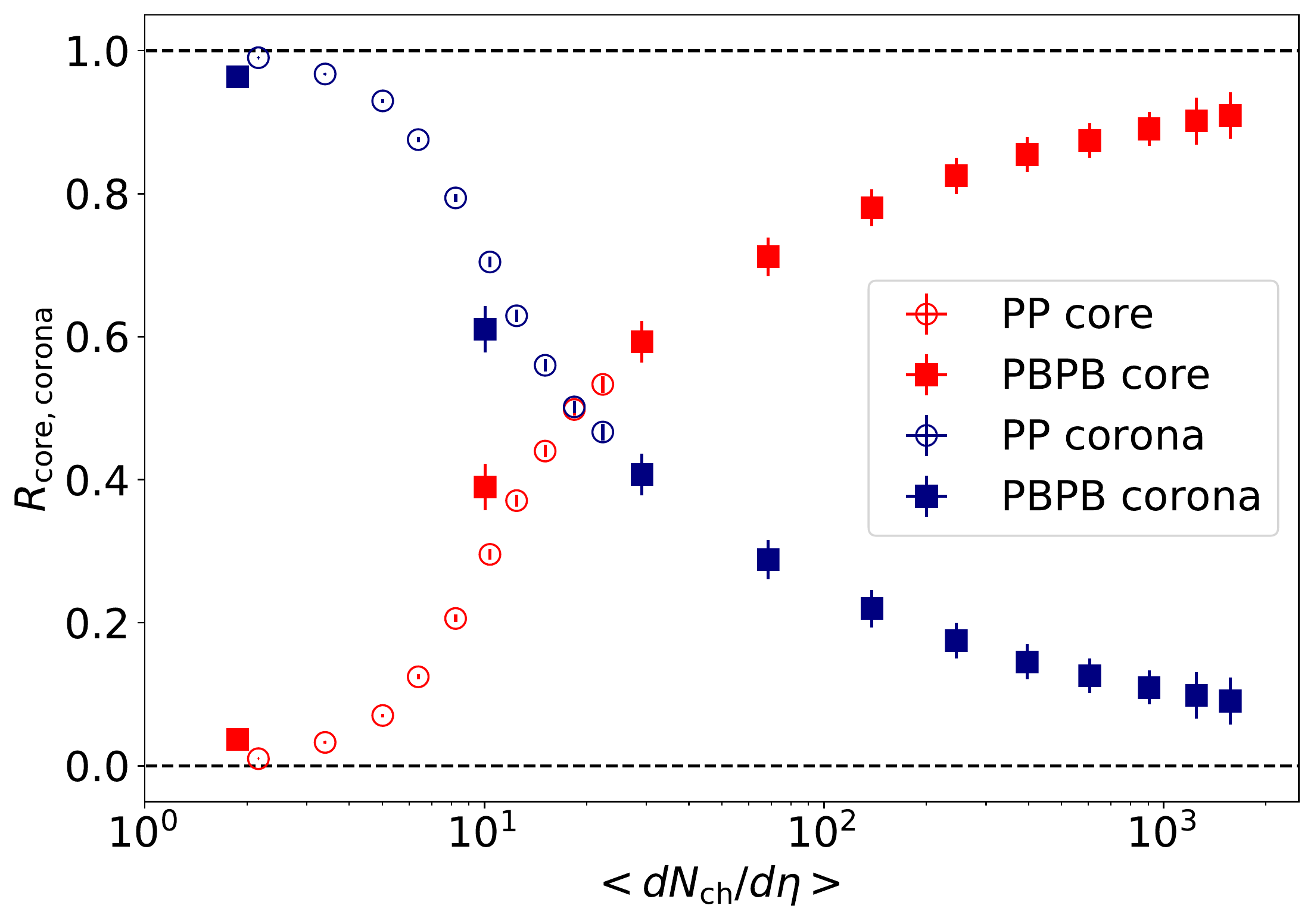}
\caption{(Color Online)
Fractions of core and corona components in the final hadron yields as functions of charged particle multiplicity at midrapidity.
Smooth behaviors of fractions of core (open red circles) and corona (open blue circles) contributions in p+p collisions at $\sqrt{s} = 7$ TeV are taken over by
those of core (closed red squares) and corona (closed blue squares) contributions in Pb+Pb collisions at $\sqrt{s_{NN}} = 2.76$ TeV, respectively.
}
\label{fig:FRACTION_CORECORONA}
\end{center}
\end{figure}

\begin{figure*}[htbp]
\includegraphics[bb=0 0 630 597, width=0.45\textwidth]{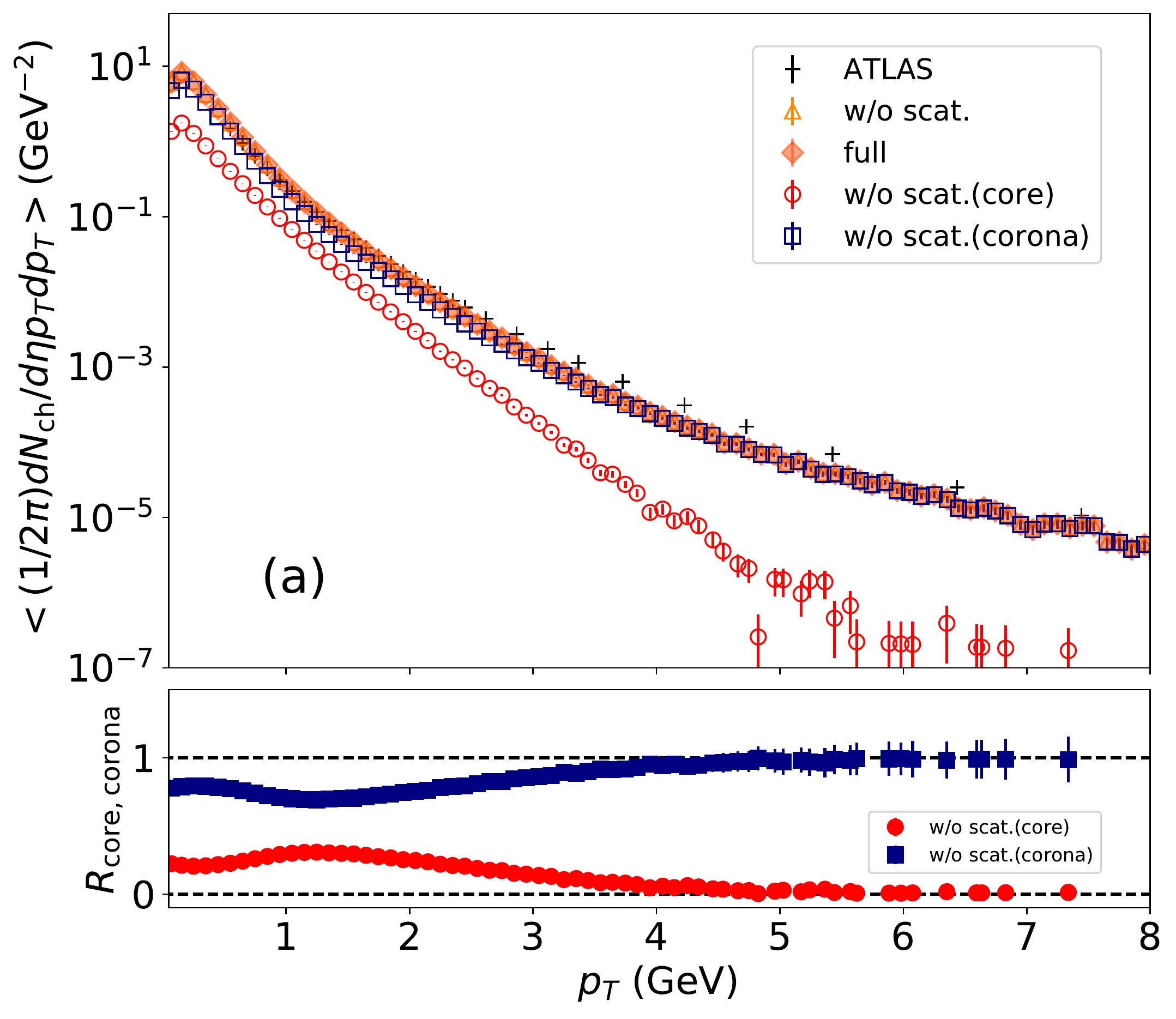}
\includegraphics[bb=0 0 630 597, width=0.45\textwidth]{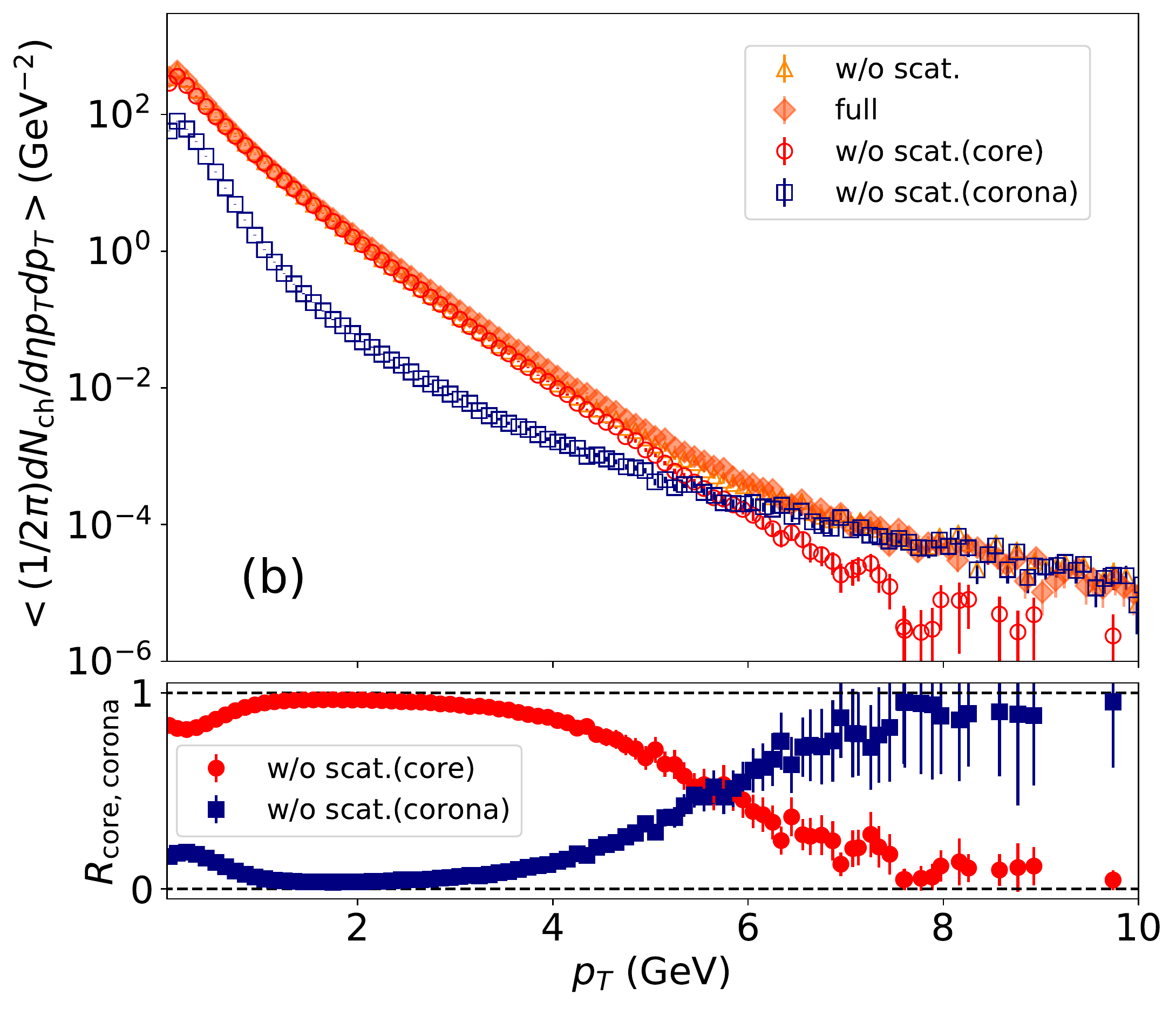}
\caption{(Color Online) (Upper) Transverse momentum spectra of charged particles at midrapidity from DCCI2 in (a) p+p collisions at $\sqrt{s} = 7$ TeV  and (b) Pb+Pb collisions at $\sqrt{s_{NN}} = 2.76$ TeV.
An event average is taken with at least one charged particle having $p_T>0.5$ GeV and $|\eta|<2.5$ in both cases.
Results from full simulations (closed orange diamonds) and simulations without hadronic rescatterings (open orange triangles) are plotted and compared with the ATLAS data (black pluses) \cite{Aad:2010ac} only in p+p collisions as a reference.
Results from core (open red diamonds) and from corona (open blue squares) are also plotted for simulations without hadronic rescatterings.
(Lower) Corresponding fractions of core (red circles) and corona (blue squares) components in the final hadron without hadronic rescatterings are shown as functions of transverse momentum. 
}
\label{fig:PTSPECTRA_PP_PBPB}
\end{figure*}

Figure~\ref{fig:FRACTION_CORECORONA} shows the fractions of core and corona components in p+p and Pb+Pb collisions simultaneously, which are identical to the results in the lower panels in Fig.~\ref{fig:MULTIPLICITY_PP_PBPB} but as functions of charged particle multiplicity at midrapidity.
Smooth crossover from corona dominance to core dominance
appears along multiplicity from p+p to Pb+Pb collisions.
The dominant contribution flips at $\langle dN_{\mathrm{ch}}/d\eta \rangle_{|\eta|<0.5} \sim 18$.
These results clearly demonstrate that the fractions of contribution from core and corona components are 
scaled with charged particle multiplicity in DCCI2, 
regardless of differences in the system size or collision energy
between p+p and Pb+Pb collisions.
Here we emphasize that, interestingly, the fraction of corona still remains $\sim 10$\% at the most central events in Pb+Pb collisions.
This also implies that both core and corona components should be implemented even in dynamical modeling of high-energy heavy-ion collisions towards precision studies on properties of QCD matter.

\subsection{Transverse momentum dependence of core and corona contribution}
\label{sec:TRANSVERSE_MOMENTUM_CORECORONA}

It is also interesting to compare the sizes of core and corona contributions in transverse momentum $p_T$ spectra of the final state particles. 
According to the implementation of the core--corona picture by
Eq.~\refbra{eq:four-momentum-deposition}, initial low momentum partons are likely to generate QGP fluids and expected to contribute largely in the low-$p_T$ region. 
Meanwhile, high momentum particles are likely to traverse vacuum or fluids as mostly keeping their initial momentum and supposed to dominate the high-$p_T$ region. 

Upper panels of Fig.~\ref{fig:PTSPECTRA_PP_PBPB} show the charged particle $p_T$ spectra at midrapidity $|\eta|<2.5$ in (a) p+p and (b) Pb+Pb collisions.
The kinematic cuts and event selections are the same as the ones used in the ATLAS experimental results \cite{Aad:2010ac}.
Event average is made with at least one charged particle having $p_T>0.5$ GeV and $|\eta|<2.5$ in both p+p and Pb+Pb collisions, which can be regarded as almost minimum-bias events.
Again, a comparison between results obtained from full simulations and ones from simulations without hadronic rescatterings is made here.
Each contribution from core and corona components to the final hadrons from simulations without hadronic rescatterings is shown as well.
In both p+p and Pb+Pb collisions, the $p_T$ spectra of final hadrons without hadronic rescatterings are represented as sums of contributions of
core and corona components over the whole $p_T$ regions.
One also sees that the effect of hadronic rescatterings on the $p_T$ spectra of charged particles is almost absent in both p+p and Pb+Pb collisions.
Since the charged particles are mainly composed of charged pions, their $p_T$ spectra are relatively insensitive to hadronic rescatterings.

\begin{figure*}[htbp]
\begin{center}
\includegraphics[bb=0 0 527 619, width=0.45\textwidth]{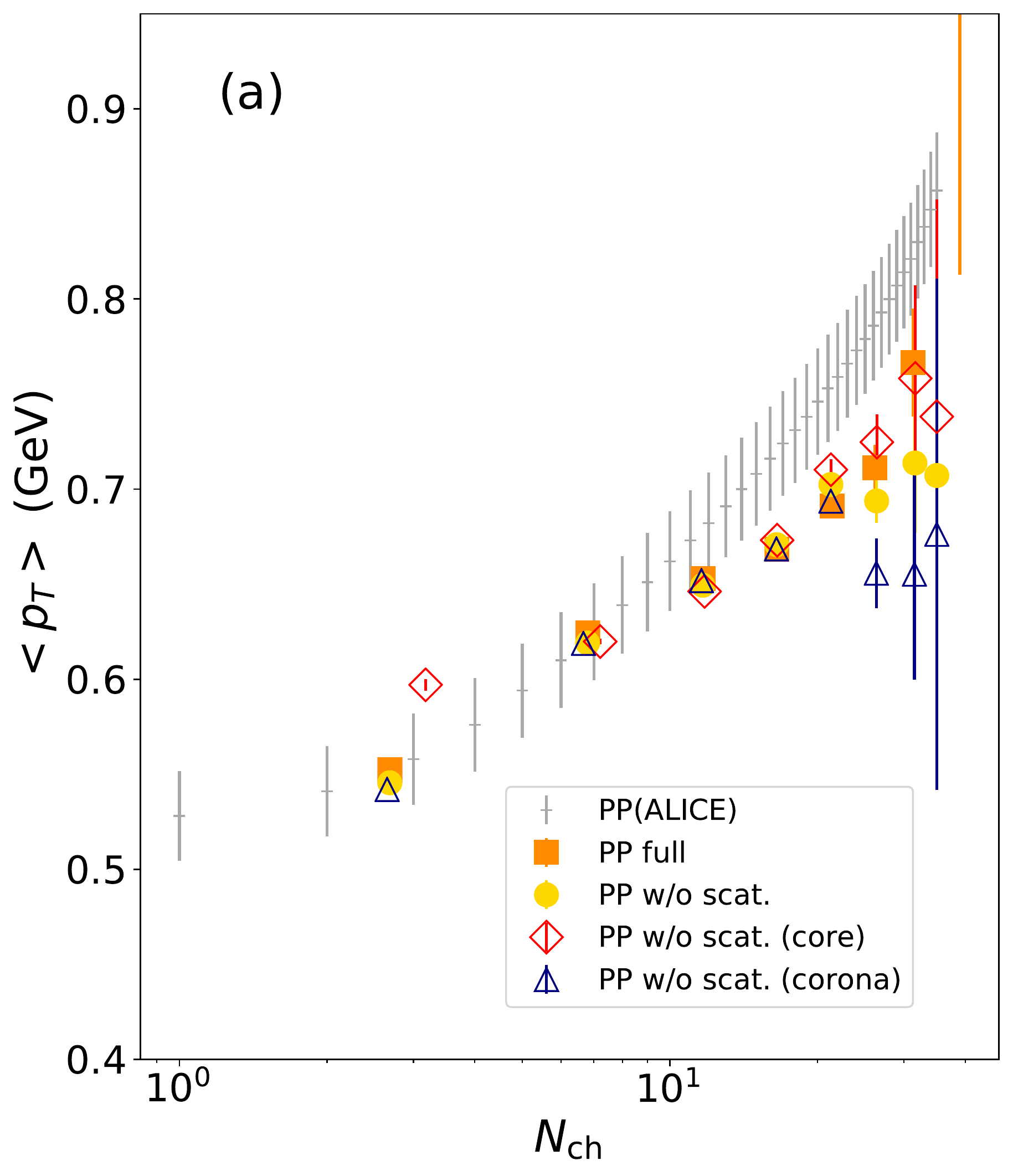}
\includegraphics[bb=0 0 527 619, width=0.45\textwidth]{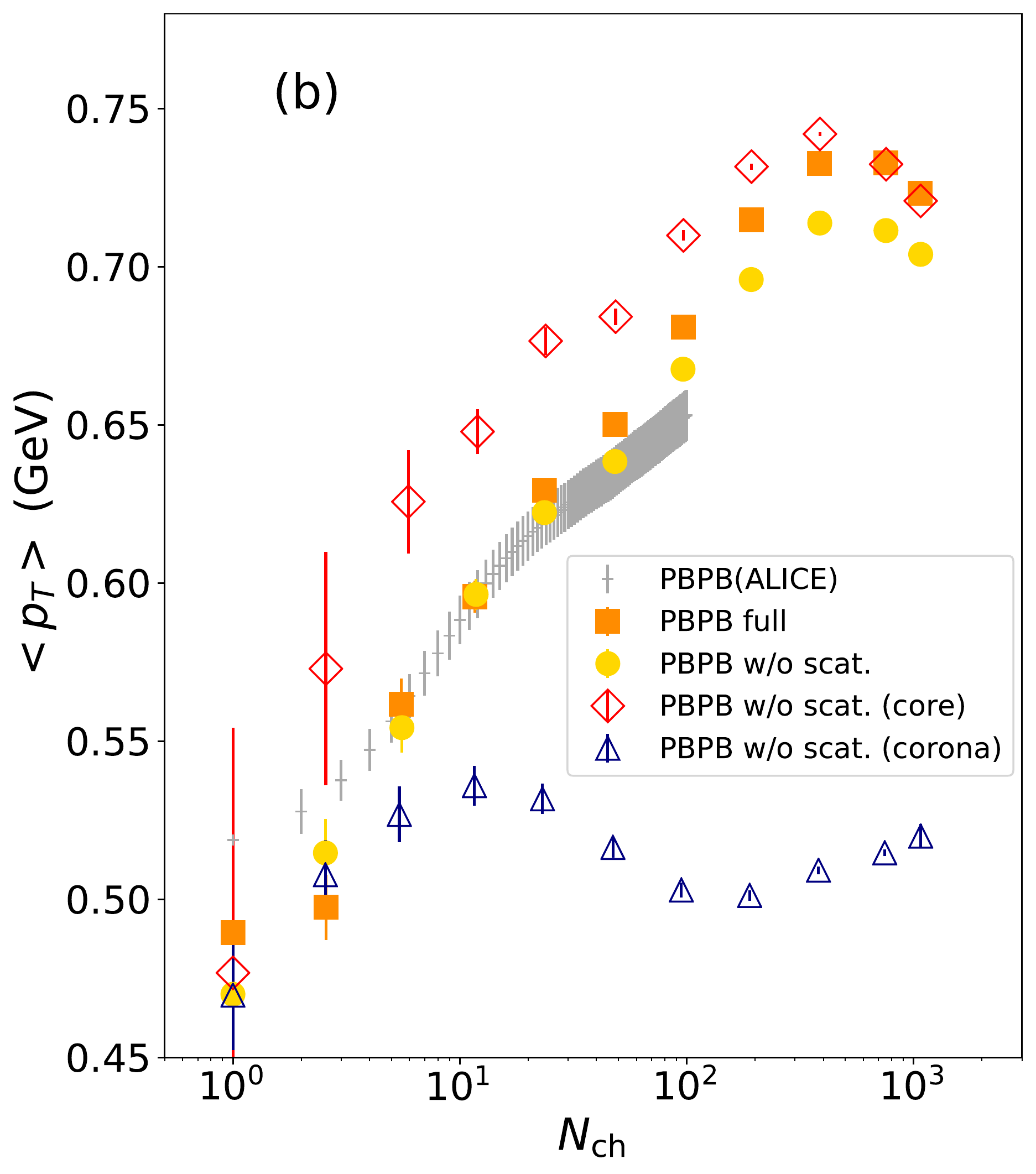}
\caption{(Color Online)
Mean transverse momentum of charged particles as a function of the number of charged particles in (a) p+p collisions at $\sqrt{s} = 7$ TeV and (b) Pb+Pb collisions at $\sqrt{s_{NN}} = 2.76$ TeV compared with the ALICE experimental data \cite{Abelev:2013bla}
(gray pluses). Results from full simulations (closed orange squares), from simulations without hadronic rescatterings (closed yellow circles), from core components (open red diamonds), and from corona components (open blue triangles) are shown for comparisons.
}
\label{fig:MEANPTMULTI_PP_PBPB_CORECORONA}
\end{center}
\end{figure*}

The corresponding lower panels of Fig.~\ref{fig:PTSPECTRA_PP_PBPB} show the fractions of core and corona components for final hadrons without hadronic rescatterings as functions of $p_T$.
As an overall tendency, the dominance of the corona components at high $p_T$ regions is seen in both p+p and Pb+Pb collisions, which is exactly what we expect from the core--corona picture in the momentum space encoded in Eq.~\refbra{eq:four-momentum-deposition}.
In p+p collisions, the contribution from the core components reaches $R_{\mathrm{core}} \sim 0.3$ around $p_T \sim 1.0$-$1.5$ GeV
and the contribution from the corona components is almost dominant 
over the whole $p_T$ range. 
On the other hand, the contribution from core components is dominant in low $p_T$ regions in Pb+Pb collisions, 
while the dominant contribution is flipped to the corona components at $p_T \sim 5.5$ GeV towards high $p_T$ regions. 
Remarkably, 
only within $0.7 \lesssim p_{T} \lesssim 3.6$ GeV,
the core components highly dominate, $R_{\mathrm{core}} \gtrsim 0.9$.
The existence of corona components should be considered below $\sim 0.7$ GeV and above $\sim3.6$ GeV even in minimum-bias events.

In particular, there is a small peak in the fraction of corona components with $R_{\mathrm{corona}} \sim 0.2$ at most in $p_T\lesssim1$ GeV.
This contribution originates mainly from a feed-down from fragmentation of strings including surviving partons during the dynamical initialization stage.
This is a consequence of the dynamical core--corona initialization against initially generated partons. Thus there should be a kind of ``redshift" of the $p_T$ spectrum due to energy loss of traversing partons which contribute as corona components in the soft region.
As we emphasized in Introduction, 
in the core--corona picture, this result exactly illustrates ``soft-from-corona" that there exists a non-negligible contribution of non-equilibrated corona components in low $p_{T}$ region.
Therefore, 
in order to properly extract transport coefficients of the QGP fluids from, for example, an analysis of flow observables, hydrodynamic results should be corrected with corona components.
We demonstrate this correction within DCCI2 in the next subsection.

\subsection{Correction from corona to flow observable}
\label{sec:CORECORONA_CONTRIBUTION}
As we discussed in Secs.~\ref{subsection:PARAMETER_DETERMINATION} and \ref{sec:TRANSVERSE_MOMENTUM_CORECORONA}, 
both core and corona contributions appear over a wide range of multiplicity.
Moreover, each component contributes as a function of $p_T$ in a nontrivial way.
To investigate how the effects of the interplay between core and corona components appear on observable, we first analyze 
the mean transverse momentum $\langle p_T \rangle$ of charged particles at midrapidity as a function of
the number of charged hadrons generated at midrapidity $N_{\mathrm{ch}}$ in p+p and Pb+Pb collisions.

Figure~\ref{fig:MEANPTMULTI_PP_PBPB_CORECORONA} shows the mean transverse momentum $\langle p_T\rangle$ of charged particles as a function of charged particle multiplicity $N_{\mathrm{ch}}$ in 
(a) p+p collisions at $\sqrt{s} = 7 \ \mathrm{TeV}$ and (b) Pb+Pb collisions at $\sqrt{s_{NN}} = 2.76 \ \mathrm{TeV}$.
Charged particles with $0.15<p_T<10.0$ GeV and $|\eta|<0.3$ are used for evaluation of $\langle p_T\rangle$,
while $N_{\mathrm{ch}}$ is obtained by counting charged particles with $|\eta|<0.3$ (without $p_T$ cut), which is the same kinematic range used in Ref.~\cite{Abelev:2013bla}.

For p+p collisions in Fig.~\ref{fig:MEANPTMULTI_PP_PBPB_CORECORONA} (a), our result from DCCI2 qualitatively describes the steep enhancement of $\langle p_T\rangle$ along $N_{\mathrm{ch}}$ observed in the ALICE experimental data \cite{Abelev:2013bla}.
Almost no significant difference is seen between results from full simulations and the ones without hadronic rescatterings.
This means that the effect of hadronic rescatterings on $\langle p_T\rangle$ of charged particles is almost negligible due to a small number of final hadrons in p+p collisions.
One also sees that 
the core and corona components show small difference of $\langle p_T \rangle$ below $N_{\mathrm{ch}} \sim 20$.
This is because, as seen in Fig.~\ref{fig:PTSPECTRA_PP_PBPB} (a),
there is no large difference for the slopes of $p_T$ spectrum of the core and corona components in low $p_T$ regions while the particle productions in the region would contribute to $\langle p_T \rangle$ significantly.

\begin{figure*}[htbp]
\begin{center}
\includegraphics[bb=0 0 539 611, width=0.45\textwidth]{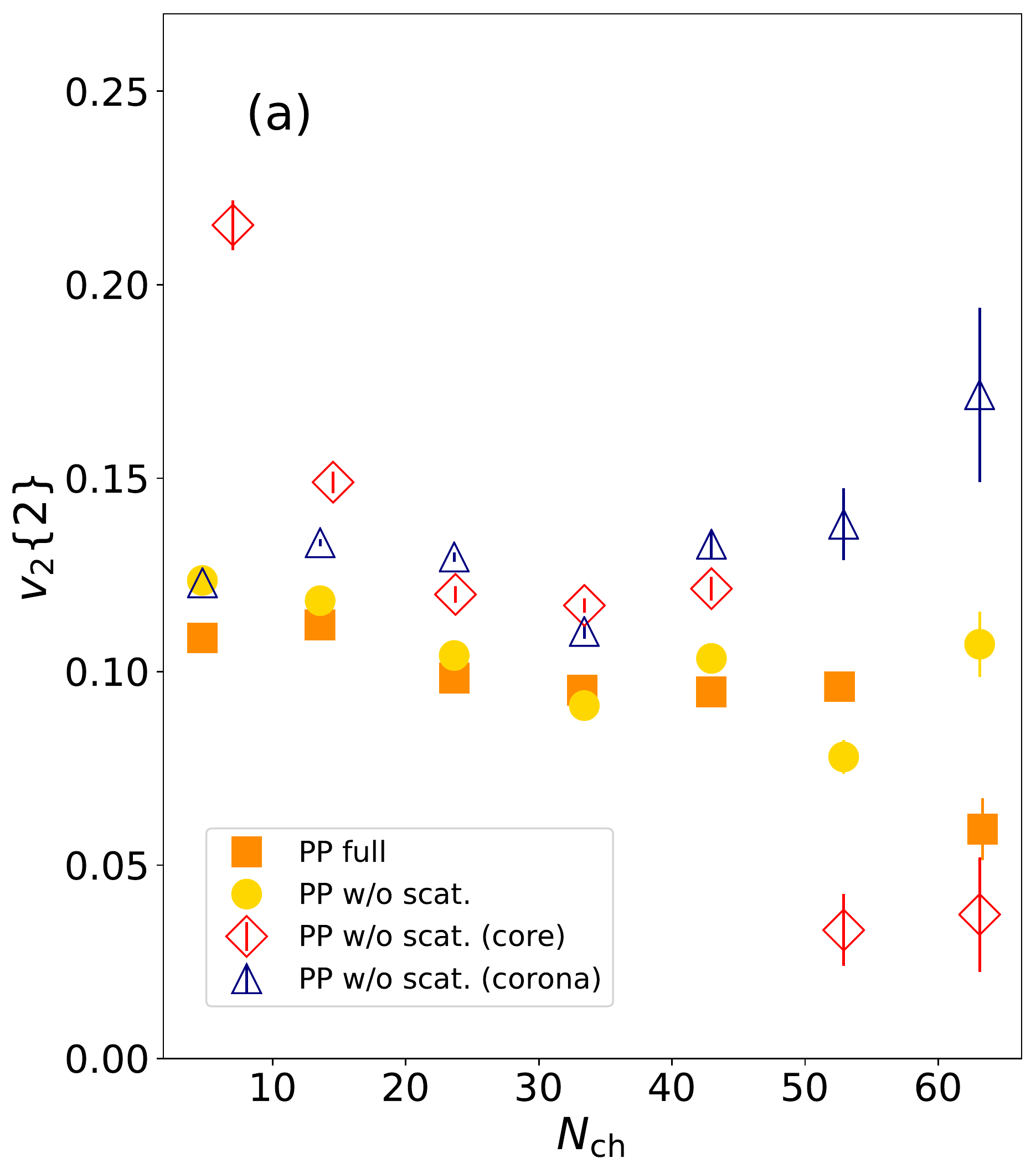}
\includegraphics[bb=0 0 539 611, width=0.45\textwidth]{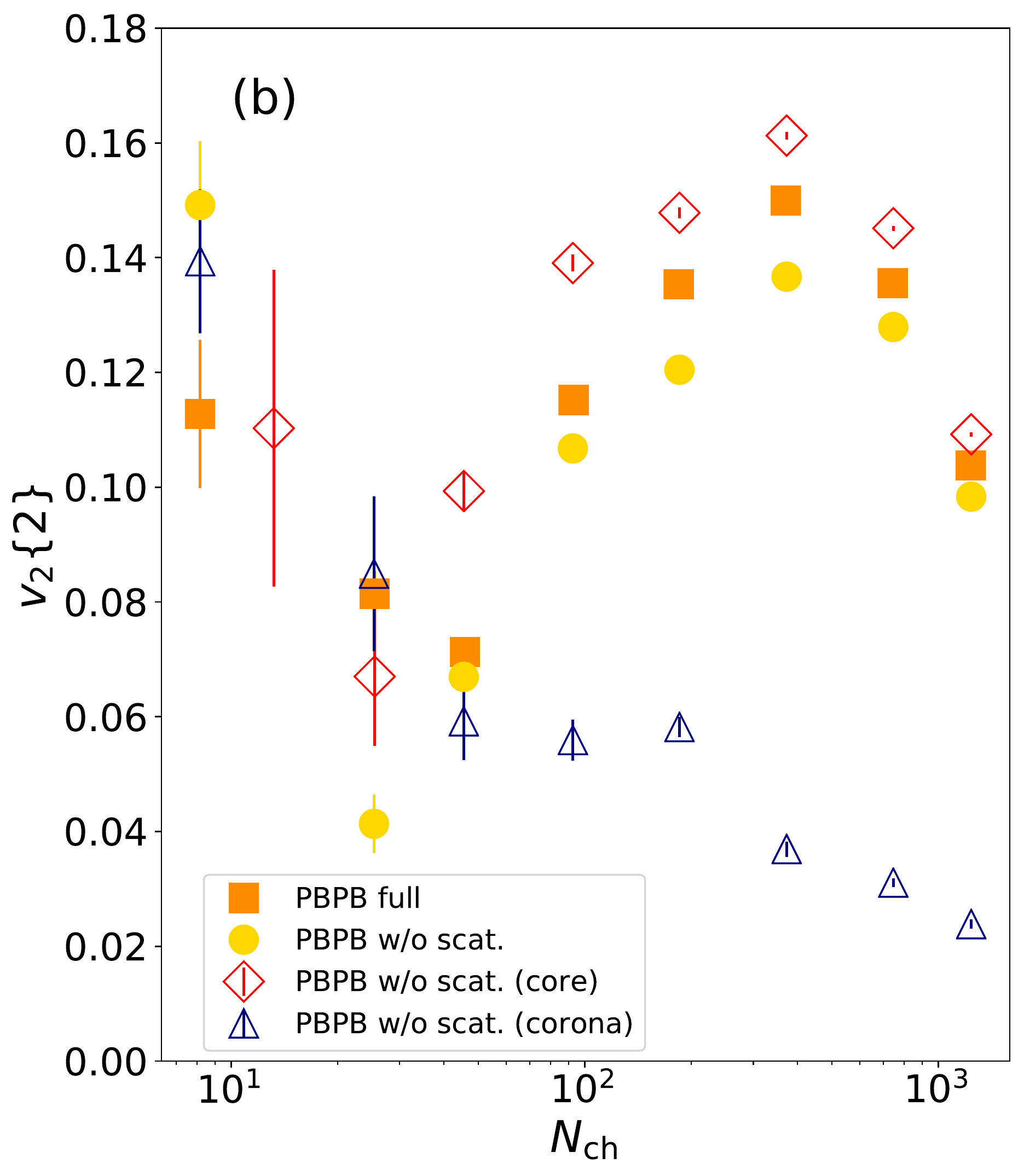}
\caption{(Color Online)
Second order of anisotropic flow coefficient obtained from two-particle correlation for charged hadrons as a function of the number of produced charged particles in (a) p+p collisions at $\sqrt{s} = 7 \ \mathrm{TeV}$ and (b) Pb+Pb collisions at $\sqrt{s_{NN}} = 2.76 \ \mathrm{TeV}$. Results from full simulations (orange squares) and simulations without hadronic rescatterings (yellow diamonds) are shown with closed symbols. While results from core (red diamonds) and corona (blue triangles) components from simulations without hadronic rescatterings are plotted with open symbols.
}
\label{fig:V22MULTI_PP_PBPB_CORECORONA}
\end{center}
\end{figure*}

For Pb+Pb collisions presented in Fig.~\ref{fig:MEANPTMULTI_PP_PBPB_CORECORONA} (b), our results from full simulations with DCCI2 reasonably describe the experimental data within the range of experimental data. 
A slight difference is seen between results from full simulations and the ones from simulations without hadronic rescatterings:
Mean transverse momentum is slightly enhanced due to hadronic rescatterings and the effect becomes relatively clear as increasing $N_{\mathrm{ch}}$. 
On the other hand, the large difference is seen between the results from core and corona components.
The core components show larger $\langle p_T\rangle$ while the corona components show smaller values for almost the entire $N_{\mathrm{ch}}$. 
The larger $\langle p_T\rangle$ from core components originates from the flatter slope of $p_T$ spectrum,
while 
the smaller $\langle p_T\rangle$ from corona components originates from the steeper slope of $p_T$ spectrum in the low $p_T$ region seen in Fig.~\ref{fig:PTSPECTRA_PP_PBPB} (b).
The difference between the results without hadronic rescatterings and the ones from core components exactly exhibits there exists the sizable \textit{correction from non-thermalized matter} to the results obtained purely from hydrodynamics.
The correction is found to be visible for the entire  $N_{\mathrm{ch}}$ and to be $\sim 5$-$11\%$ in $N_{\mathrm{ch}}\lesssim 200$.
Therefore the ``soft-from-corona" components are the key to precisely reproduce the multiplicity dependence of the mean transverse momentum
 \footnote{
It is discussed that the centrality dependence of {$\langle p_T\rangle$} is well described by hydrodynamic simulations introducing the finite bulk viscosity \cite{Ryu:2015vwa}.
While a recent Bayesian analysis supports the zero-consistent bulk viscosity by analyzing $p_T$-differential observables \cite{Nijs:2020ors}.
Both of them, however, still failed to reproduce pion $p_T$ spectra below $\sim 0.3$ GeV (see, \textit{e.g.}, Fig.~3 in Ref.~\cite{Ryu:2015vwa} and Fig.~20 in Ref.~\cite{Nijs:2020roc}. Another related discussion is in Ref.~\cite{Guillen:2020nul}), which would result in overestimation of {$\langle p_T\rangle$}.
The discrepancy between hydrodynamic results and experimental data in this low $p_T$ region becomes larger as going to peripheral collisions or small colliding systems \cite{Nijs:2020roc}. Therefore the deviation between the model and the data in the low $p_T$ region could be filled with the corona components and would improve the description of {$\langle p_T\rangle$}. 
}.

Figure~\ref{fig:V22MULTI_PP_PBPB_CORECORONA} shows the second-order anisotropic flow coefficient of charged particles obtained from two-particle cumulants, $\vtwtw$, as a function of $N_{\mathrm{ch}}$ in (a) p+p collisions at $\sqrt{s}=7$ TeV and (b) Pb+Pb collisions at $\sqrt{s_{NN}} = 2.76$ TeV. 
Kinematic cuts for both $\vtwtw$ and $N_{\mathrm{ch}}$ are $0.2<p_T<3.0$ GeV and $|\eta|<0.8$ as used in Ref.~\cite{Acharya:2019vdf}.
It should be mentioned that insufficient statistics with DCCI2 simulations do not allow us to have a pseudorapidity gap of charged hadron pairs $|\Delta \eta| > 1.4$ in the $\vtwtw$ analysis unlike in the ALICE analysis.
This is the reason why we do not compare our results with the experimental data in this paper.
We leave quantitative discussion by comparing with experimental data for future work. 

For the results in p+p collisions,
$\vtwtw$ obtained from both core and corona components is larger than that from simulations without hadronic rescatterings in $10 \lesssim N_{\mathrm{ch}} \lesssim 50$.
This suggests that the event plane angle of core components might be different from that of corona components, which dilutes $\vtwtw$ of core and corona components with each other.

For the results in Pb+Pb collisions, 
the $\vtwtw$ from full simulations reaches a maximum value at $N_{\mathrm{ch}}\sim 400$, which is similar to the tendency observed in experimental data \cite{Acharya:2019vdf}. 
From a comparison between the results with and without hadronic rescatterings, one can tell that a slight enhancement of $\vtwtw$ comes from generation of elliptic flow in the late hadronic rescattering stage \cite{Hirano:2005xf,Hirano:2007ei,Takeuchi:2015ana}.
Here again, one can see the correction from corona components in the comparison between the core result and the inclusive result in the case without hadronic rescatterings. 
The correction from corona components is found to be $\sim 15$-$38\%$ below $N_{\mathrm{ch}} \sim 370$, which originates from the small peak seen at very low $p_T$ region in the $p_T$ spectra in Fig.~\ref{fig:PTSPECTRA_PP_PBPB} (b) \footnote{The leftmost point of the contribution from core components is slightly shifted to large $N_{\mathrm{ch}}$ since there are some events  in which one cannot calculate two-particle cumulants due to less than two charged particles from the core parts are measured in a given kinematic window in this $N_{\mathrm{ch}}$ bin. Therefore the event average of $N_{\mathrm{ch}}$ for core components is biased to larger $N_{\mathrm{ch}}$.}.
This suggests that one would need to incorporate corona components in hydrodynamic frameworks to extract transport coefficients from comparisons with experimental data.

In both p+p and Pb+Pb results,
there are two factors that would give a finite anisotropy in corona components, which are color reconnection and feed-down from surviving partons.
The color reconnection effect implemented in default \pythia8 and \pythia8 Angantyr can arise collectivity \cite{Bierlich:2018lbp}.
With the color reconnection, dense color strings formed due to multiparton interactions interact with each other and eventually induce flow-like behavior of final hadrons.
Its effect can be enhanced due to more multiparton interactions in initial parton generation with DCCI2 compared to default \pythia8 and \pythia8 Angantyr. 
The detailed discussion on multiparton interactions in initial parton generation is made in Sec.~\ref{subsection:Evolution_of_transverse_energy}.
Under the dynamical core--corona initialization, partons originating from hard scatterings and emitted in the back-to-back directions tend to survive.
In contrast, soft partons, which originate from multiparton interactions and are randomly directed, tend to be converted into fluids.
Since the low $p_T$ charged hadrons come from such surviving partons through string  fragmentation, $\vtwtw$ of corona components could reflect that of their parents.
As a result, the corona components show larger anisotropy compared to results from the default \pythia8 \cite{Bierlich:2018lbp} and \pythia8 Angantyr.

\subsection{Multiplicity dependence of mean transverse mass}
\label{sec:MTSCALING}
The fraction of core components to total hadronic productions increases along charged particle multiplicity as shown in Fig.~\ref{fig:FRACTION_CORECORONA}.
Since the effects of radial flow are expected to be more pronounced as increasing fraction of core components,
we analyze the mean transverse mass for various hadrons in high- and low-multiplicity p+p and Pb+Pb events and see its mass dependence.
It has been empirically known that $m_T$ spectra in small colliding systems exhibit the $m_T$ scaling, \textit{i.e.}, the slope of $m_T$ spectra being independent of the rest mass of hadrons \cite{Guettler:1976ce,Guettler:1976fc}. 
Here, $m_T=\sqrt{m^2 + p_T^2}$ is the transverse mass and $m$ is the rest mass of the hadron. 
In contrast, in heavy-ion collisions, the slope parameter increases with $m$
and, as a result, the $m_T$ scaling is violated, which is regarded as a sign of the existence of radial flow generated \cite{Bearden:1996dd,Xu:2001zj}.
Thus whether \textit{radial expansion exists in small colliding systems} due possibly to the QGP formation can be explored through the empirical scaling behavior and its violation in the mean transverse mass. 

We take two event classes, high-multiplicity ($0$-$10\%$) and low-multiplicity ($50$-$100\%$) events, in p+p and in Pb+Pb collisions \footnote{Due to the lack of statistics, we simply divide events into these classes regardless of collision system.}.
The multiplicity or centrality classification is performed in the same way as the one used in Fig.~\ref{fig:MULTIPLICITY_PP_PBPB}.
In the following, we analyze the mean transverse mass
of charged pions ($\pi^+$ and $\pi^-$), charged kaons ($K^+$ and $K^-$), protons ($p$ and $\bar{p}$), phi mesons ($\phi$), lambdas ($\Lambda$ and $\bar{\Lambda}$), cascade baryons ($\Xi^-$ and $\bar{\Xi}^+$), and omega baryons ($\Omega^-$ and $\bar{\Omega}^+$), in $|\eta|<0.5$ without $p_T$ cut.

\begin{figure*}[htbp]
\begin{center}
\includegraphics[bb=0 0 468 507, width=0.35\textwidth]{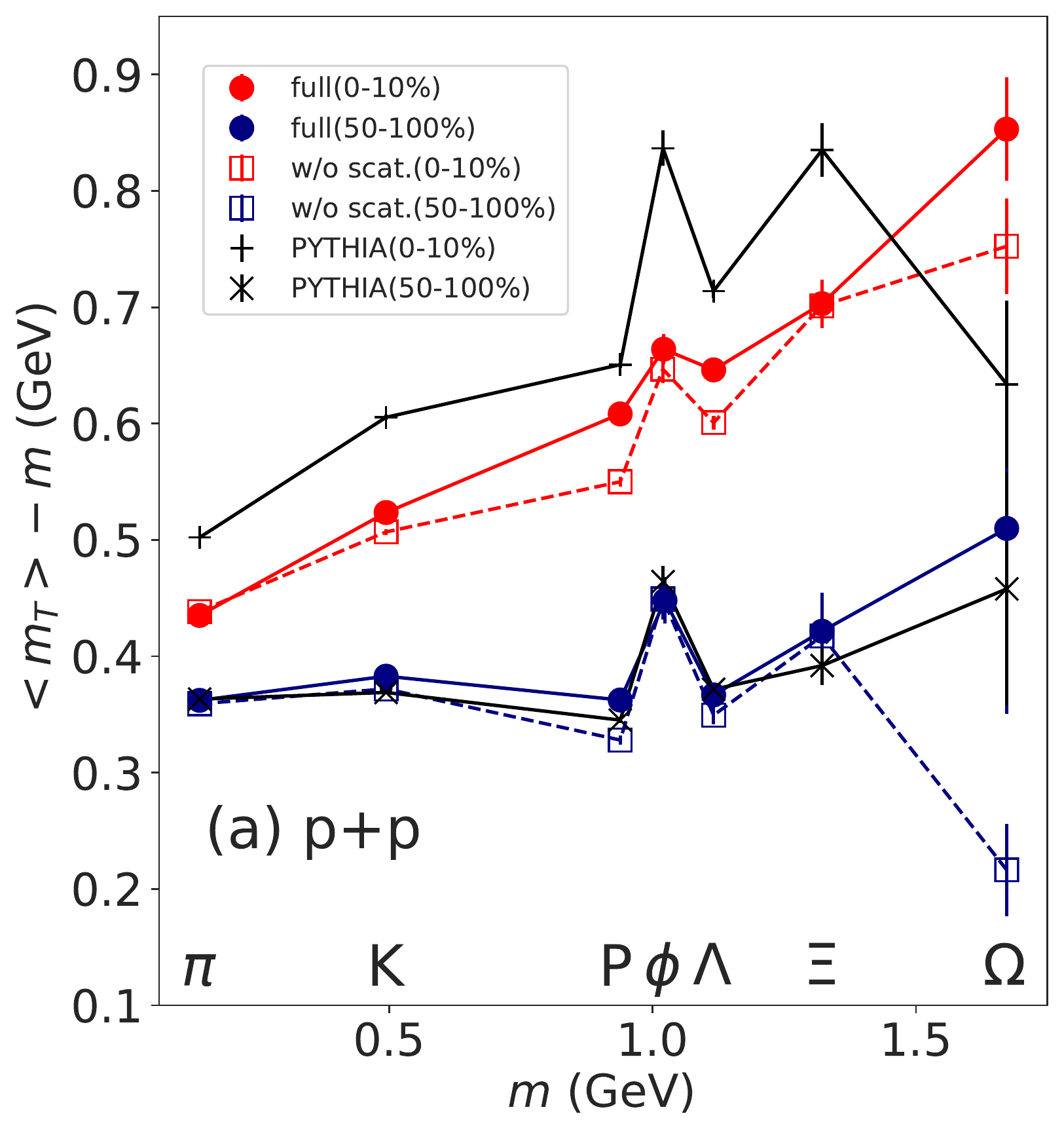}
\hspace{1.5cm}
\includegraphics[bb=0 0 468 507, width=0.35\textwidth]{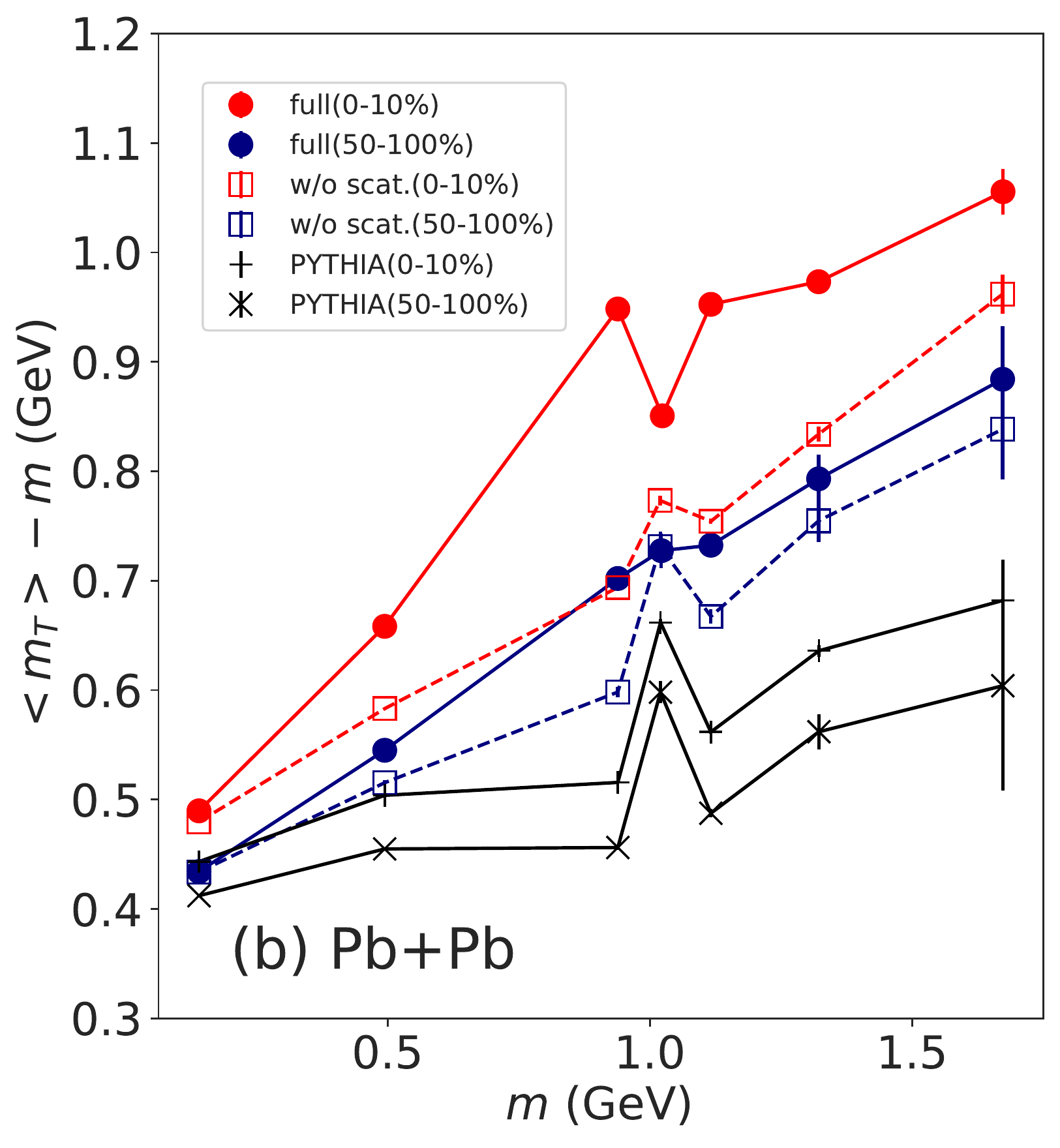}
\includegraphics[bb=0 0 468 507, width=0.35\textwidth]{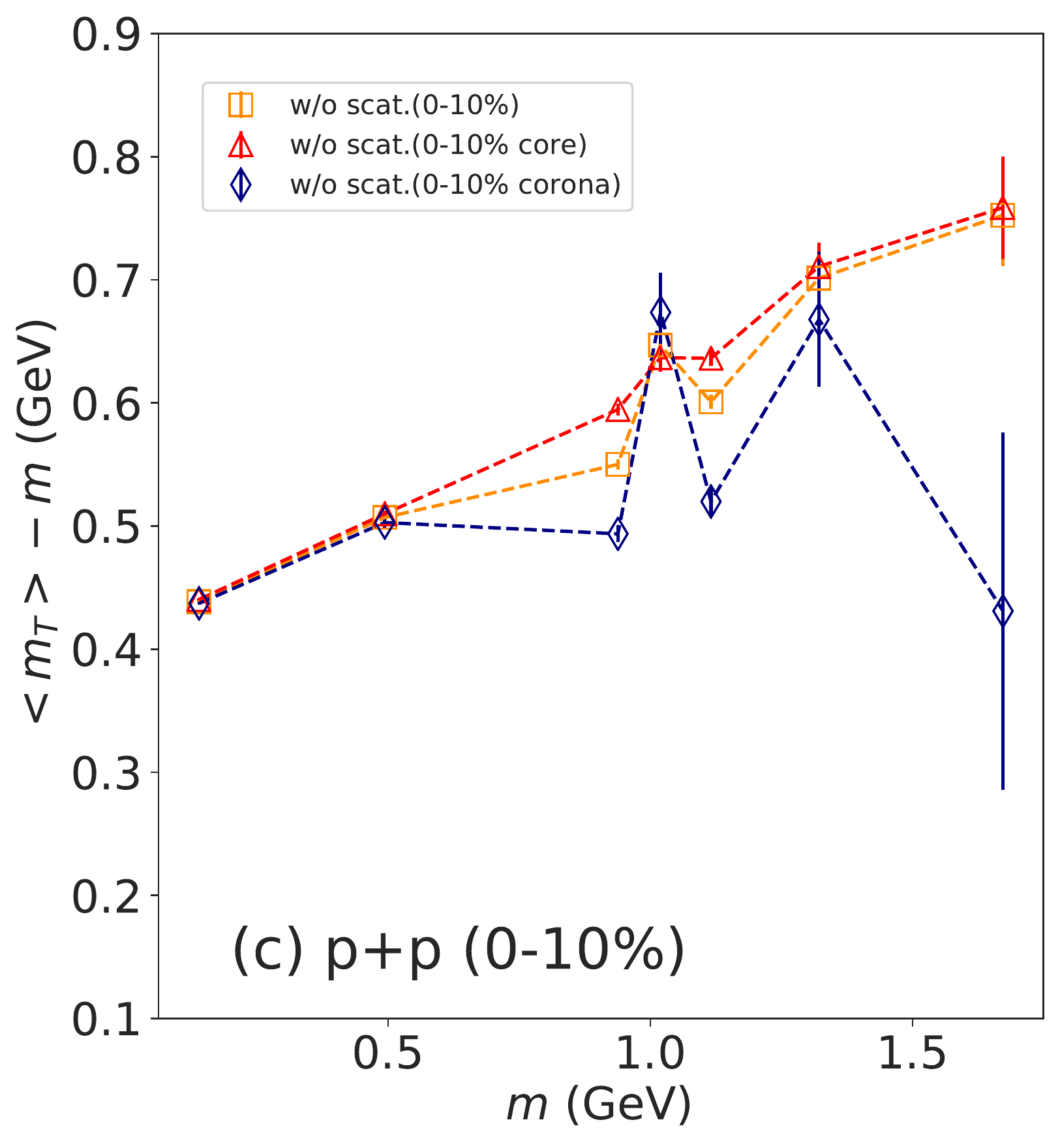}
\hspace{1.5cm}
\includegraphics[bb=0 0 468 507, width=0.35\textwidth]{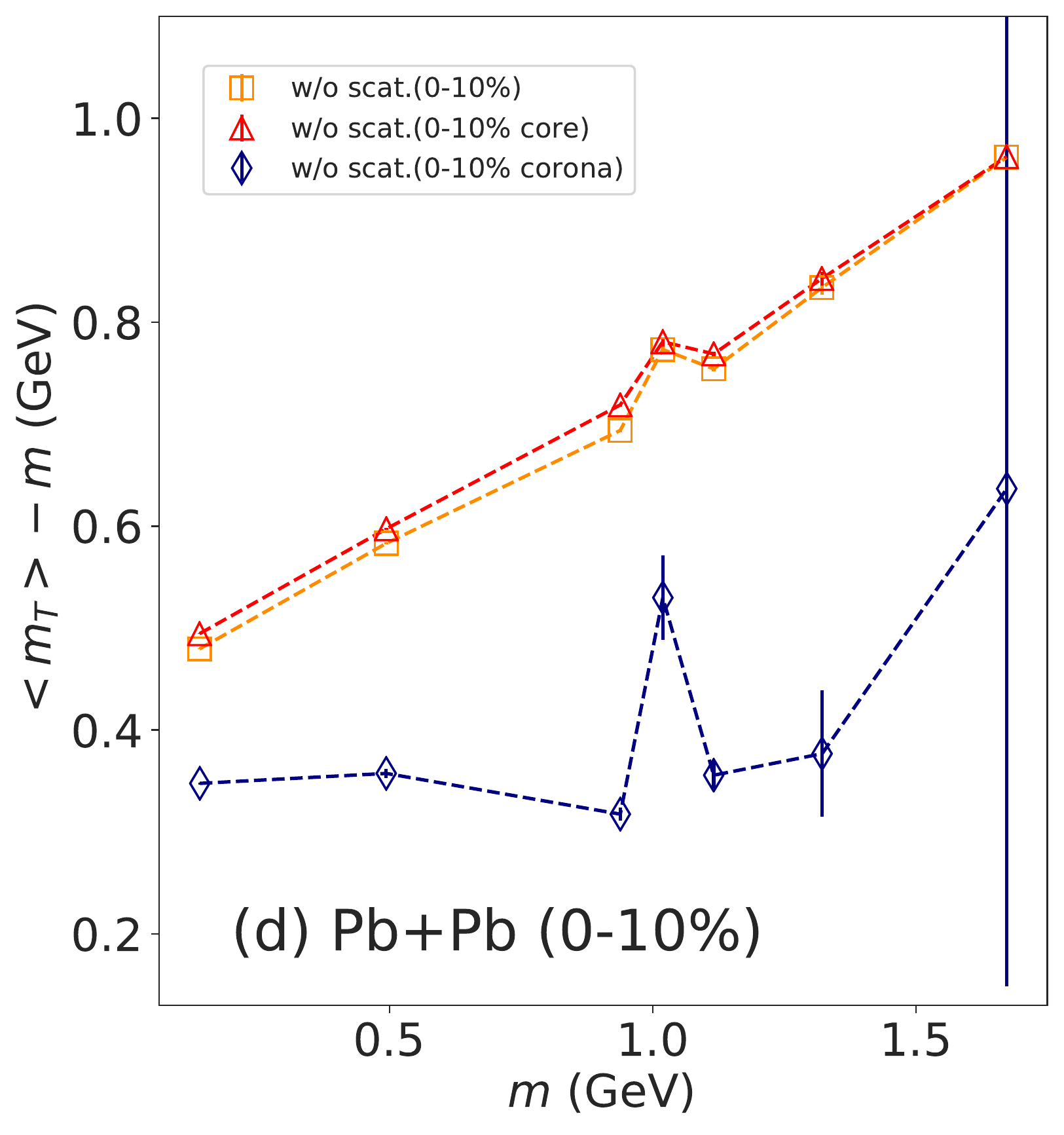}
\includegraphics[bb=0 0 468 507, width=0.35\textwidth]{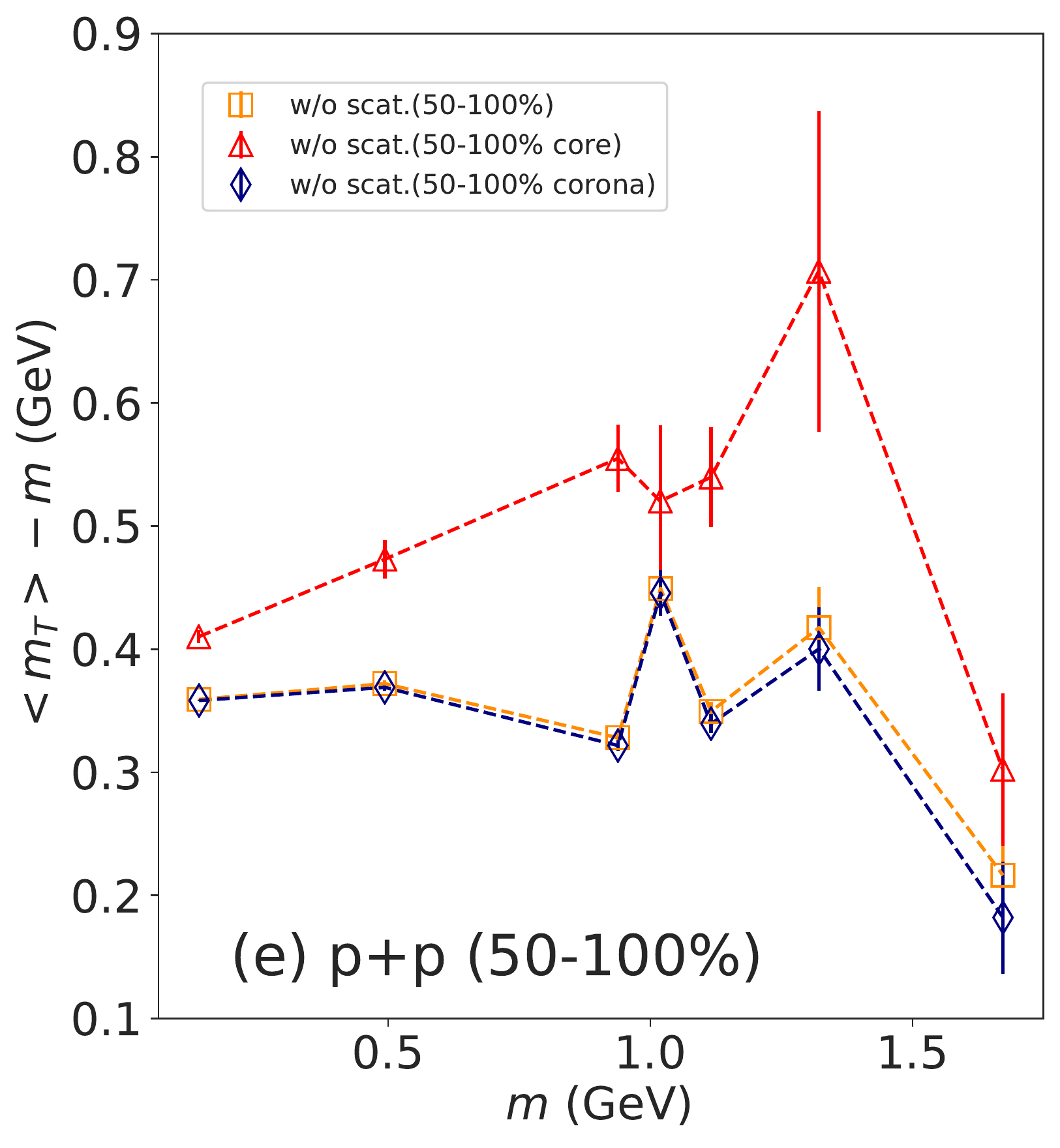}
\hspace{1.5cm}
\includegraphics[bb=0 0 468 507, width=0.35\textwidth]{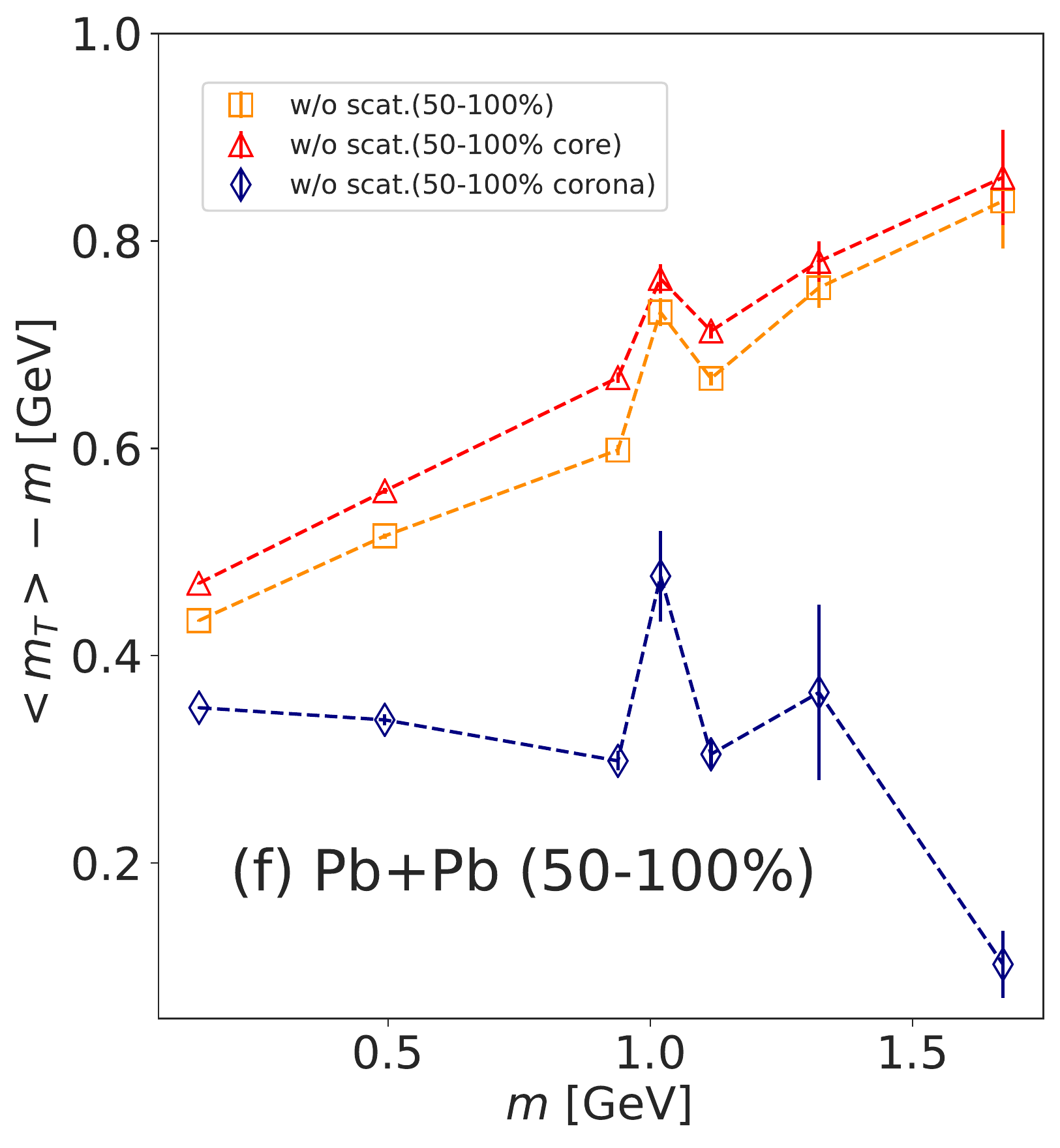}
\caption{(Color Online) Mean transverse mass, $\langle m_T \rangle -m$, as a function of rest mass of hadrons, $m$, from DCCI2 in (a) p+p collisions at $\sqrt{s} = 7 \ \mathrm{TeV}$ and (b) Pb+Pb collisions at $\sqrt{s_{NN}} = 2.76 \ \mathrm{TeV}$. A comparison of the results from full simulations (closed symbols connected with solid lines) and the ones from simulations without hadronic rescatterings (open symbols connected with dashed lines) is made. Results of high-multiplicity events ($0$-$10\%$, red) and of low-multiplicity events  ($50$-$100\%$, blue)  are shown to see the effects of the fraction of core components. 
The result from \pythia8 and \pythia8 Angantyr (black symbols) with default parameters including color reconnection is plotted in p+p and Pb+Pb collisions, respectively, as references.
Corresponding contributions of core and corona components in $0$-$10\%$ multiplicity and centrality classes in (c) p+p and in (d) Pb+Pb collisions, respectively.
Corresponding contribution of core and corona components in $50$-$100\%$ multiplicity and centrality classes in (e) p+p and in (f) Pb+Pb collisions, respectively. }
\label{fig:MTSCALING_PP_PBPB}
\end{center}
\end{figure*}

Figure~\ref{fig:MTSCALING_PP_PBPB} (a) shows the mean transverse mass, $\langle m_T \rangle -m$, as a function of the rest mass of hadrons, $m$, in high-multiplicity ($0$-$10\%$) and low-multiplicity ($50$-$100\%$) p+p collisions at $\sqrt{s}= 7$ TeV.
To pin down the effect of hadronic rescatterings, the results from full simulations and simulations without hadronic rescatterings are compared with each other.
The result from \pythia8 with the default settings including color reconnection is also plotted as a reference. 
Overall, $\langle m_T \rangle-m$ in high-multiplicity events ($0$-$10\%$) tends to exhibit an almost linear increase with increasing $m$ except for phi mesons.
On the other hand, such a clear mass dependence is not seen in low-multiplicity events ($50$-$100\%$), which is consistent with the $m_T$ scaling.
It should also be mentioned that the violation of linear increase of phi mesons in core components appears after resonance decays against direct hadrons (not shown). 
An apparent flow-like linear mass dependence is seen in results from \pythia8 in high-multiplicity events as well, which is due to the color reconnection \cite{Ortiz:2013yxa}.
The almost linear increasing behavior in DCCI2 is caused by both radial flow for core components from hydrodynamic expansion and color reconnection for corona components obtained with \pythia8.
As a result, the results from both DCCI2 and \pythia8 have similar tendencies.
Therefore, it is difficult to discriminate each effect by merely seeing the mean transverse mass.
The effect of hadronic rescatterings is almost absent for pions. This comes from an interplay between small $pdV$ work in the late hadronic rescattering stage and approximate conservation of pion number~\cite{Hirano:2005wx}. 
The small effects of hadronic rescatterings are seen for phi mesons and omega baryons because they do not form resonances in scattering with pions unlike other hadrons~\cite{Hirano:2007ei,Takeuchi:2015ana}.

Figure~\ref{fig:MTSCALING_PP_PBPB} (b) shows the mean transverse mass as a function of hadron rest mass for $0$-$10\%$ and $50$-$100\%$ centrality classes in Pb+Pb collisions at $\sqrt{s_{NN}} = 2.76$ TeV. 
The almost linear increasing behavior of $\langle m_T \rangle-m$ appears even in $50$-$100\%$ centrality class as one can expect from the centrality dependence of the fraction of core components in Fig.~\ref{fig:MULTIPLICITY_PP_PBPB} (b).
The larger enhancement of the mean transverse mass due to hadronic rescatterings, in particular, for protons is seen in high-multiplicity events in comparison with the low-multiplicity events.
This is a manifestation of the famous ``pion wind" in the late rescattering stage \cite{Hung:1997du,Bleicher:1999pu,Bratkovskaya:2000qy,Bass:2000ib}.

Figure~\ref{fig:MTSCALING_PP_PBPB} (c) shows each contribution of core and corona components to the final result without hadronic rescatterings in 0-10\% multiplicity class in p+p collisions. The inclusive result here is identical to the one shown as the result without hadronic rescatterings in Fig.~\ref{fig:MTSCALING_PP_PBPB} (a).
The difference between results of core and corona components is seen in protons, lambdas, and omega baryons.
The linear mass ordering of $\langle m_T \rangle-m$ from core components is slightly diluted 
for protons and lambdas in the inclusive result
due to the sizable contribution of corona components.
In contrast, the core result and the inclusive result are almost on top of each other since the contribution of corona components for omega baryons is smaller in 0-10\% multiplicity class compared to other particle species.

Figure~\ref{fig:MTSCALING_PP_PBPB} (d) shows each contribution of core and corona components to the final result without hadronic rescatterings in 0-10\% centrality class in Pb+Pb collisions. The linear increase except phi meson is seen very clearly for the core component, and the increase rate is more than the one from p+p results shown in Fig.~\ref{fig:MTSCALING_PP_PBPB} (c).

Figure \ref{fig:MTSCALING_PP_PBPB} (e) shows the same variable with Fig.~\ref{fig:MTSCALING_PP_PBPB} (c) but in 50-100\% multiplicity class in p+p collisions.
Since the fraction of the core components is less than 10\% in this range of multiplicity class as shown in Fig.~\ref{fig:MULTIPLICITY_PP_PBPB} (a), the final result and the result from corona components are almost top of each other showing no significant dependence on hadron rest mass. 

Figure~\ref{fig:MTSCALING_PP_PBPB} (f) shows the same variables with Fig.~\ref{fig:MTSCALING_PP_PBPB} (d) but in 50-100\% centrality class in Pb+Pb collisions. According to Fig.~\ref{fig:MULTIPLICITY_PP_PBPB} (b), the fraction of core components shows $R_{\mathrm{core}}\sim 0.8$ to $\sim 0.9$ in this centrality range. Eventually the result of core components is found to be slightly diluted by corona components.

\subsection{Evolution of transverse energy}
\label{subsection:Evolution_of_transverse_energy}

As shown in Fig.~\ref{fig:MULTIPLICITY_PP_PBPB} in Sec.~\ref{subsection:PARAMETER_DETERMINATION}, we reproduced centrality dependence of charged particle multiplicity in p+p and Pb+Pb collisions within DCCI2.
Although the default \pythia8 (or Angantyr model in heavy-ion modes) works reasonably well, reproduction of multiplicity within DCCI2 can be attained only after the considerable change of a parameter $p_{\mathrm{T0Ref}}$ from its default value as mentioned in Sec.~\ref{subsection:Parameter_set_in_DCCI2}.
A nontriviality in DCCI2 stems from different competing mechanisms of how the transverse energy changes during the evolution of the system.
In this subsection, we discuss the effects of string formation/fragmentation and longitudinal $pdV$ work on the transverse energy and explain why we needed to change this parameter in DCCI2.

The transverse energy per unit rapidity $dE_{T}/d\eta$ is a basic observable in high-energy nuclear collisions and contains rich information on the dynamics of an entire stage of the reactions.
The transverse energy changes mainly in the initial and the expansion stages of the reactions.
In the initial stage, the two energetic hadrons and/or nuclei form color flux tubes between them as they pass through each other.
The chromo-electric and magnetic fields in the color flux tubes possess the energy originating from the kinetic energy of colliding hadrons or nuclei.
The decays of color flux tubes into partons and subsequent rescatterings among them are supposed to lead to the QGP formation \cite{Kajantie:1985jh,Gatoff:1987uf,Eskola:1992bd}.
Thus, how much energy is deposited in the reaction region is a fundamental problem of the QGP formation and depends on the initial dynamics of high-energy nuclear collisions.
On the other hand, in the expansion stage, the $pdV$ work associated with the longitudinal expansion after the QGP formation reduces the energy produced in the initial reaction region \cite{Gyulassy:1983ub,Ruuskanen:1984wv}.
The amount of reduced energy is sensitive to viscosity and other transport properties of the QGP~\cite{Gyulassy:1997ib}.
Therefore $dE_{T}/d\eta$ can be a good measure to scrutinize modeling in the initial and the expansion stages of the reactions.

In \pythia8, partons are first generated through hard scatterings and then, together with partons from initial and final state radiations, form hadron strings which eventually fragment into hadrons.
The transverse energy per unit rapidity of final hadrons is always larger than that of initially generated partons around midrapidity.
To understand this enhancement around midrapidity, suppose a hadronic string formed from a di-quark in the forward beam rapidity region and a quark in the backward beam rapidity region as an extreme case. 
Although the partons lie only around beam rapidity regions and the transverse energy of them vanishes around midrapidity, that string fragments into hadrons almost uniformly in rapidity space.
Thus, the emergence of the transverse energy at midrapidity is a consequence of the formation of a color string between such partons around beam rapidity.
Since parameters in \pythia8 are so tuned to reproduce the final hadron spectra, the initial parameters are highly correlated with parameters in the fragmentation as a whole.
Therefore the default parameter set should not be used if the subsequent hydrodynamic evolution, which reduces the transverse energy from its initial value of generated partons, is incorporated in DCCI2.

\begin{figure*}[htpb]
\begin{center}
\includegraphics[bb=0 0 1175 336, width=0.95\textwidth]{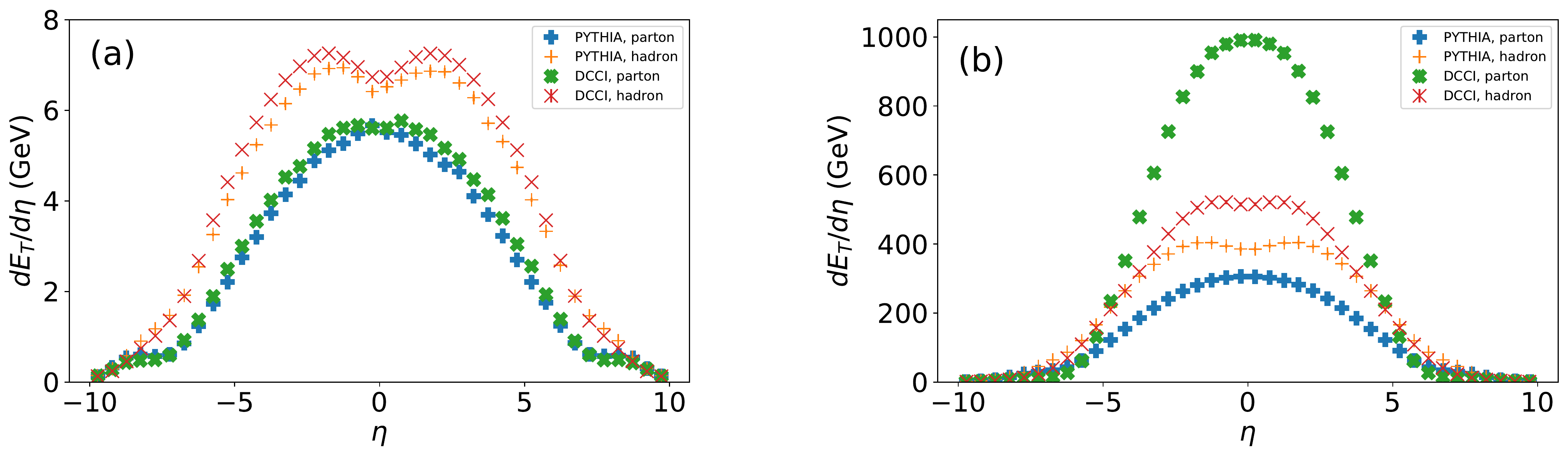}
\caption{(Color Online)
Pseudo-rapidity distribution of transverse energy in (a) INEL$>0$ p+p collisions at $\sqrt{s} = 7$ TeV 
and (b) minimum-bias Pb+Pb collisions at $\sqrt{s_{NN}} = 2.76$ TeV from DCCI2 and \pythia8.
A comparison of transverse energy distribution between parton and hadron levels is made for results from DCCI2 and \pythia8.
Results from the parton level in DCCI2 (green crosses) and \pythia8 (blue pluses) and the ones from the hadron level (red crosses) in DCCI2 and \pythia8 (orange pluses) are shown for comparison. For Pb+Pb collisions, \pythia8 Angantyr is used to obtain the results.
}
\label{fig:SHIFTOFDETDETA}
\end{center}
\end{figure*}

Figure \ref{fig:SHIFTOFDETDETA} shows $dE_{T}/d\eta$
of the initial partons before the string hadronization or the dynamical initialization, and that of the hadrons in the final state from both \pythia8 and DCCI2 in minimum-bias (a) p+p collisions at $\sqrt{s} = 7 $ TeV and (b) Pb+Pb collisions at $\sqrt{s_{NN}} = 2.76$ TeV.
For the results from \pythia8 in Fig.~\ref{fig:SHIFTOFDETDETA} (a), 
the transverse energy per unit pseudorapidity, $dE_{T}/d\eta$, of the final hadrons is always larger than that of the initial partons in the whole rapidity region except around the beam rapidity.
Since the final hadron yield is dominated by the corona components in p+p collisions in DCCI2, these results are almost identical with the ones from \pythia8.
For the results from \pythia8 Angantyr in Fig.~\ref{fig:SHIFTOFDETDETA} (b),
the behavior of the transverse energy in Pb+Pb collisions is again the same as in p+p collisions.
In contrast, to obtain the same amount of the transverse energy in the final state in DCCI2, the transverse energy must be deposited initially $\sim 3$ times as large as that in \pythia8 Angantyr at midrapidity to reconcile the reduction of transverse energy due to $pdV$ work.
This is exactly possible by considerably decreasing the parameter $p_{\mathrm{T0Ref}}$.

The parameter $p_{\mathrm{T0Ref}}$ regulates infrared divergence of the QCD cross section, can be interpreted as a parameter $p_{\perp\mathrm{min}}$ to separate soft from hard scales \footnote{In actual simulations in \pythia, a parameter $p_{\mathrm{T0Ref}}$ provides a scale to make a smooth turnoff of hard scattering rather than the sharp separation \cite{Sjostrand:2017cdm}.}, and controls the number of multiparton interactions in \pythia\ \cite{Sjostrand:1987su,Sjostrand:2017cdm}.
The smaller the separation scale $p_{\perp\mathrm{min}}$ is, the larger the number of multiparton interaction $\langle n_{\mathrm{MPI}}(p_{\perp\mathrm{min}}) \rangle = \sigma_{2\rightarrow 2}(p_{\perp\mathrm{min}})/\sigma_{\mathrm{nd}} $ is. 
Here $\sigma_{2 \rightarrow 2}$ and $\sigma_{\mathrm{nd}}$ are the perturbative QCD $2\rightarrow 2$ cross section and the inelastic non-diffractive cross section, respectively.
By increasing $\langle n_{\mathrm{MPI}}(p_{\perp\mathrm{min}}) \rangle$ as decreasing $p_{\mathrm{T0Ref}}$, initial partons are generated more and bring the sufficient amount of transverse energy in the final hadron state as shown in Fig.~\ref{fig:SHIFTOFDETDETA} (b). 
In this work, we use $p_{\mathrm{T0Ref}}=$1.8 and 0.9 for p+p and Pb+Pb collisions as mentioned in Sec.~\ref{subsection:PARAMETER_DETERMINATION}, which are smaller than the default values,
2.28 and 2.0 for
{\texttt{MultipartonInteractions:pT0Ref}} and {\texttt{SpaceShower:pT0Ref}} 
in \pythia.

So far, we have found that 
hydrodynamics and string fragmentation have different evolution of transverse energy. Thus, the multiplicity of final hadrons from such a two-component model is sensitive to a fraction of each component in a system.
To, at least, reproduce multiplicity in DCCI2, we need to change the parameter $p_{\mathrm{T0Ref}}$ from its default value.
However, as we mentioned at the beginning of this section, the other parameter $\sigma_0$ in Eq.~\refbra{eq:four-momentum-deposition} 
has a non-trivial correlation with $p_{\mathrm{T0Ref}}$,
which means that we need to tune both the parameters at the same time. 
Suppose that one firstly tries to reproduce multiplicity by tuning $p_{\mathrm{T0Ref}}$. 
Since a small $p_{\mathrm{T0Ref}}$ gives a rise to the number of multiparton interactions and 
deposited transverse energy is enhanced, it affects to enhance final hadron multiplicity.
On the other hand,
since the number of initial partons produced in midrapidity increases,
this causes more fluidization in dynamical core--corona initialization.
Once a fraction of the core is enhanced, the multiplicity of final hadrons can also decrease since the initial transverse energy deposited in midrapidity region is used for $pdV$ work.
As a result of competition between these effects, multiplicity cannot linearly enhance or decrease by decreasing or increasing $p_{\mathrm{T0Ref}}$. 
Secondly, suppose that one tries to reproduce particle yield ratios as functions of multiplicity by tuning $\sigma_0$. 
Since changing $\sigma_0$ means changing a fraction of core and corona, final multiplicity is easily altered, too.
This is the reason why we need to fix both parameters by taking into account multiplicity and particle yield ratios at the same time.

Note that, if we made viscous corrections in the hydrodynamic evolution, the resultant change of  $p_{\mathrm{T0Ref}}$ from its default value could have been modest due to the less reduction of transverse energy \cite{Gyulassy:1997ib}, which is beyond the scope of the present paper, but which should be investigated in the future work.

A string melting version of A Multi Phase Transport (AMPT) model  \cite{Lin:2001zk,Lin:2004en} and the hydrodynamic models using it for generating initial conditions \cite{Pang:2012he, Xu:2016hmp} avoid this issue of the transverse energy in an ``ad hoc" way: The hadrons decaying from a  string are re-decomposed into their constitutive quarks and antiquarks, and then form high-energy density partonic matter. 
Although it is possible to count the energy stored along the string contrary to considering the generated partons directly, this prescription lacks gluons from melting strings.
Therefore, we do not pursue this idea in the present paper.

\section{Summary}
\label{sec:summary}
We studied the interplay between core and corona components establishing the DCCI2, which describes the dynamical aspects of core--corona picture under the dynamical initialization scheme.
To develop the DCCI2, we put an emphasis on reconciliation of open issues of dynamical models, mainly relativistic hydrodynamic models, toward a comprehensive description of a whole reaction of high-energy nuclear collisions.
One of the important achievements is to generate the initial profiles of hydrodynamics by preserving initial total energy and momentum of the collision systems.
This is achieved by adopting hydrodynamic equations with source terms on initial partons obtained from \pythia8, one of the widely accepted general-purpose Monte-Carlo event generators.
Consequently, in addition to the equilibrated matter (core) described by relativistic hydrodynamics, we also consider the existence of non-equilibrated matter (corona) through dynamical initialization with the core--corona picture.

We have updated our model from the previous work.
The updates include sophistication of four-momentum deposition of initial partons in dynamical core--corona initialization, samplings of hadrons from hypersurface of core parts (fluids) with \ISthreeD, hadronic afterburner for final hadrons from core and corona parts with a hadronic transport model \jam, and modification on color string structures in corona parts due to co-existence with core parts (fluids) in coordinate space.

Discussion on the interplay between core and corona components is made once 
fixing major parameters so that our model reasonably describes multiplicity as a function of multiplicity or centrality class and omega baryon yield ratios to charged pions as functions of multiplicity.
First we extracted the fractions of core and corona components to the final hadron yields as functions of multiplicity and centrality classes. We found that, as increasing multiplicity, the core components become dominant at $\langle dN_{\mathrm{ch}}/d\eta \rangle_{|\eta|<0.5} \sim 18$, which corresponds to about 0.95-4.7\% multiplicity class in p+p collisions at $\sqrt{s} = 7$ TeV and $\sim80\%$ centrality class in Pb+Pb collisions at $\sqrt{s_{NN}} = 2.76$ TeV.
Next, we showed the fractions of core and corona components in charged particle $p_T$ spectra. 
In minimum-bias Pb+Pb collisions, the fraction of core components is dominant below $p_T\sim 5.5$ GeV, while that of corona components is dominant above that.
Interestingly, we found that there was an enhancement in the fraction of corona contribution with $R_{\mathrm{corona}} \sim 0.2$ at most in $p_T \lesssim 1$ GeV even in minimum-bias Pb+Pb collisions. From this, the fraction of the corona contribution is anticipated to increase in peripheral collisions. 
This brings up a problem in all conventional hydrodynamic calculations in which low $p_T$ soft hadrons are regarded purely as core components.
Since the fraction of each component would exist finite for a wide range of multiplicity and, as a result, there should be interplay between them, we suggest that 
both small colliding systems and heavy-ion collisions 
should be investigated in a unified theoretical framework by incorporating both core and corona components.

To investigate the effects of co-existence of core and corona components on observables, we showed $\langle p_T \rangle$ and $v_{2}\{2\}$ as functions of $N_{\mathrm{ch}}$. 
In particular, in Pb+Pb collisions, 
we found that the finite contribution of corona components at midrapidity gives a certain correction on the results obtained purely from core components, which is described by hydrodynamics.
The correction is $\sim 5$-$11$\% for $\langle p_T \rangle$ below $N_{\mathrm{ch}}\sim 200$,
while it is 
$\sim 15$-$38$\% for $\vtwtw$ below $N_{\mathrm{ch}}\sim 370$.
The former correction leads to the reasonable agreement of $\langle p_T \rangle$ with the experimental data. 
This suggests that one might need to incorporate corona components in hydrodynamic frameworks to extract transport coefficients from comparisons with experimental data.
Finally we explored effects of radial flow based on violation of $m_T$ scaling with hadron rest mass by classifying events into high- and low-multiplicity ones.
Noteworthy, we found that it is difficult to discriminate the radial flow originated from hydrodynamics from collectivity arisen from color reconnection in \pythia8.
We also discussed evolution of transverse energy in the DCCI2.
In string fragmentation, final transverse energy is larger than initial transverse energy as producing hadrons around midrapidity. While in hydrodynamics, transverse energy just decreases from its initial value during the evolution due to the longitudinal $pdV$ work.
To obtain the same amount of transverse energy in the final state in DCCI2 with default \pythia8 Angantyr in minimum-bias Pb+Pb collisions, it is necessary to have three times larger initial transverse energy than the one of default \pythia8 Angantyr.

For more quantitative discussion on transport properties of the QGP fluids, we admit an absence of viscous corrections to fully equilibrium distribution in our analysis.
We leave this as one of our future works.
Nevertheless, we emphasize that the corrections from corona components mean the ones from ``far from'' equilibrium components which should exist nonetheless and would more significantly affect the final hadron distributions than the viscous corrections.

It is known that transverse momentum spectra solely from hydrodynamics or hybrid (hydrodynamics followed by hadronic cascade) models do not perfectly reproduce the experimental data below $p_T \sim 0.5$ GeV \cite{Abelev:2013vea},
although hydrodynamics is believed to provide a better description in the low $p_T$ region in general. 
The deviation between pure hydrodynamic results and the data in the low $p_T$ region could be filled with the corona components. 
Detailed analyses of centrality dependent particle identified $p_T$ spectra from DCCI2, which require high statistics, and its comparison with the experimental data will be made in a future publication.

With this model, we anticipate that it would be interesting to explore 
planned O+O collisions at LHC \cite{Citron:2018lsq} since the collision system can provide data around ``sweet spot'' 
in which 
the core components are to be dominant and yet corrections from corona components cannot be ignored at all 
\cite{Brewer:2021kiv}.
In addition, investigation on strangeness enhancement in forward or backward regions might give some insights into ultra high-energy cosmic ray measurements \cite{Anchordoqui:2016oxy,Baur:2019cpv}.
Incorporation of a dynamical description of kinematic and chemical pre-equilibrium stage \cite{Kurkela:2018wud,Kurkela:2018oqw} and investigation of medium modification of jets \cite{Tachibana:2019hrn,Luo:2021iay} are in our interests as well.
We leave the discussion on those topics as a future work.

\section*{Acknowledgement}
We gratefully acknowledge valuable comments from
C.~Bierlich and T.~Sj\"{o}strand on \pythia8 and \pythia8 Angantyr, D.~Evelett and M.~McNelis on iS3D, and Y.~Nara on JAM.
The work by Y.K. is supported by JSPS KAKENHI Grant Number 20J20401. 
The work by Y.T. and T.H. was partly supported by JSPS KAKENHI Grant Number JP17H02900.\\

\bibliography{inspire.bib}

\end{document}